\documentclass[12pt,preprint,natbib]{aastex}
\input epsf

\def\s2n{S^{\prime}/N}
\def\s2n{S^{\prime}/N}
\def\vecr{{\bf x}}
\def\vecdr{{\bf \Delta x}}
\def\dr{{\ell}}
\def\aq{{\langle Q \rangle}}

\def\sig{$\sigma_v$ [km/s]}
\def\alp{$\alpha$}
\def\scf{$S_0$(1pc)}
\def\mach{$M_{\rm S}$}
\def\L0{$L_0$ [pc]}

\begin{document}
\title{The Spectral Correlation Function of Molecular Clouds: A Statistical
Test for Theoretical Models}

\author{
Paolo Padoan\footnote{padoan@jpl.nasa.gov}
\affil{Jet Propulsion Laboratory, 4800 Oak Grove Drive, MS 169-506,
     California Institute of Technology, Pasadena, CA 91109-8099, USA}
Alyssa A. Goodman
\affil{Harvard University Department of Astronomy, Cambridge, MA 02138}
and Mika Juvela
\affil{Helsinki University Observatory, T\"ahtitorninm\"aki, P.O.Box 14,
SF-00014 University of Helsinki, Finland}
}

\begin{abstract}

We compute the spectral correlation function (SCF) of $^{13}$CO J=1--0
maps of molecular cloud complexes. The SCF is a power law over 
approximately an order of magnitude in spatial separation in every map.
The power law slope of the SCF, \alp, its normalization, \scf, and the
spectral line width averaged over the whole map, $\sigma_v$, are 
computed for all the observational maps. The values of \alp\ , 
\scf\ and $\sigma_v$ are combined to obtain empirical correlations to 
be used as tests for theoretical models of molecular clouds. Synthetic 
spectral maps are computed from different theoretical models, 
including solutions of the magneto-hydrodynamic (MHD) equations with 
different values of the rms Mach number of the flow and stochastic models 
with different power spectra of the velocity field. In order to compute
the radiative transfer from the MHD models it is necessary to assign the 
models a physical scale and a physical density. When these assignments
are made according to Larson type relations the best fit to the observational
correlations is obtained. Unphysical stochastic models are instead ruled out 
by the empirical correlations. MHD models with equipartition of magnetic and 
kinetic energy of turbulence do not reproduce the observational data
when their average magnetic field is oriented approximately parallel to the 
line of sight.

\end{abstract}


\section{Introduction}

Molecular clouds are observed through the emission of a number of
molecular transitions that provide a wealth of information about
their chemical composition, gas temperature and density, magnetic 
field strength, fractional ionization, structure and kinematics. 
This information is essential to our understanding of the process 
of star formation.

The interpretation of molecular emission line maps is not always unique.
The main source of uncertainty is the absence of the third spatial
dimension (along the line of sight) in the observational data. 
Statistical properties of the velocity and density distributions
along the line of sight are difficult to disentangle. Furthermore, the
components of the gas velocity on the plane of the sky are unknown.

Two dimensional images of molecular clouds are usually converted into 
three dimensional ``objects'' using the radial velocity instead of
the third spatial dimension. This method can be useful to separate
individual mass condensations from each other, since it is conceivable
that their relative velocity is larger than their internal one.
However, velocity blending or the lack of well defined condensations
along the line of sight may cause significant uncertainties
(Issa, MacLaren \& Wolfendale 1990; Adler \& Roberts 1992; 
Ballesteros--Paredes, Vazquez--Semadeni \& Scalo 1999; Pichardo et al. 
2000; Ostriker, Stone \& Gammie 2001; Lazarian et al. 2001; 
Ballesteros--Paredes \& Mac Low 2002).

\nocite{Ostriker+2001} \nocite{Lazarian+2001}

Due to the difficulty of a direct interpretation of the observational
data, a ``forward approach'' that starts from a rather general theoretical 
model and synthesizes its observational properties can be more instructive. 
Different models may sometimes satisfy the same set of 
observational constraints, but they should also provide guidance for
further observational studies that could help select the correct model.

Ideally, numerical models to be compared with observed spectral line
data cubes should be based on the numerical solutions of the MHD 
equations, in the regime of highly super--sonic turbulence, and 
on radiative transfer calculations. In some works, stochastic fields 
are used instead of the 
solution of the MHD equations and in most studies the radiative transfer 
calculation is omitted, in favor of density--weighted velocity profiles. 
The first large synthetic spectral maps of molecular transitions 
computed by solving the non--LTE radiative transfer through the density 
and velocity data cubes obtained as the numerical solution of the MHD 
equations were presented by Padoan et al. (1998), based on Juvela's
radiative transfer code (Juvela 1997), and were used in a number
of works (e.g. Padoan et al. 1999, 2000, 2001). Another new radiative transfer
code has also been used more recently to generate synthetic spectral maps
from MHD simulations (Ossenkopf 2002).

\nocite{Padoan+98cat}

A number of statistical methods have been proposed to compare
numerical models of turbulence with large spectral maps of
molecular clouds (see for example Scalo 1984; Kleiner \& Dickman 
1985, 1987; Stutzki \& Gusten 1990; Gill \& Henriksen 1990; Houlahan 
\& Scalo 1992; Hobson 1992; Langer, Wilson, \& Anderson 1993; Williams, 
De~Geus \& Blitz 1994; Miesch \& Bally 1994; Miesch \& Scalo 1995; Lis 
et al. 1996; Blitz \& Williams 1997; Heyer \& Schloerb 1997; Stutzki 
et al. 1998; Miesch, Scalo \& Bally 1999; 
Falgarone et al. 1994; Padoan et al. 1999; Mac Low \& Ossenkopf 2000; 
Bensch, Stutzki \& Ossenkopf 2001).

\nocite{Scalo84} \nocite{Kleiner+Dickman85} \nocite{Kleiner+Dickman87}
\nocite{Stutzki+Gusten90} \nocite{Gill+Henriksen90} 
\nocite{Houlahan+Scalo92} \nocite{Hobson92} \nocite{Langer+93} 
\nocite{Williams+94} \nocite{Miesch+Bally94} \nocite{Miesch+Scalo95}
\nocite{Lis+96} \nocite{Blitz+Williams97} \nocite{Heyer+Schloerb97}
\nocite{Stutzki+98} \nocite{Miesch+99} 
\nocite{Falgarone+94} \nocite{Padoan+99per} \nocite{MacLow+Ossenkopf2000} 
\nocite{Bensch+2001}

In this work we apply the spectral correlation function (SCF) method,
proposed by Rosolowsky et al. (1999) and further developed in Padoan, 
Rosolowsky \& Goodman (2001), to a number of observational and synthetic
spectral maps. We show that the slope and normalization of the SCF of 
observational maps correlate with the spectral line width. Theoretical 
models of molecular clouds should therefore yield synthetic spectral 
maps reproducing such correlations, but not all of them can. 

\nocite{Rosolowsky+99} \nocite{Padoan+2001SCF}

In the next section we briefly define the SCF, and in \S~3 we present
the observational data used in this work. The computation of the
theoretical models and synthetic spectral maps is presented in \S~4.
Results from numerical models are compared with the observational 
data in \S~5 and are discussed in \S~6. Conclusions are drawn in \S~7.

\section{The SCF Method}

The Spectral Correlation Function (SCF) measures the spatial correlation 
of spectral line profiles within a spectral map. It is sensitive to the 
properties of both the gas mass distribution and the gas velocity field 
(Rosolowsky et al. 1999; Padoan, Rosolowsky \& Goodman 2001; Padoan et 
al. 2001; Ballesteros--Paredes, Vazquez--Semadeni \& Goodman 2002).

Let $T(\vecr,v)$ be the antenna temperature as a function of velocity 
channel $v$ at map position $\vecr$. The SCF for spectra with spatial 
separation $\dr$ is: 
\begin{equation}
S_0(\dr)=\left\langle \frac{S_0(\vecr,\dr)}{S_{0,{\rm N}}(\vecr)} \right\rangle _{\vecr},
\label{1}
\end{equation}
where the average is computed over all map positions $\vecr$. 
$S_0(\vecr,\dr)$ is the SCF uncorrected for the effects of noise,
\begin{equation}
S_0(\vecr,\dr)=\left\langle 1- \sqrt{\frac{\Sigma_v[T(\vecr,v)-T(\vecr+\vecdr,v)]^2}
{\Sigma_vT(\vecr,v)^2+\Sigma_vT(\vecr+\vecdr,v)^2}} \right\rangle _{\vecdr},
\label{2}
\end{equation}
where the average is limited to separation vectors $\vecdr$ with 
$|\vecdr|=\dr$, and $S_{0,{\rm N}}(\vecr)$ is the SCF due to 
noise alone,
\begin{equation}
S_{0,{\rm N}}(\vecr)=1-\frac{1}{Q(\vecr)},
\label{3}
\end{equation}
and $Q(\vecr)$ is the ``spectrum quality'' (see discussion in Padoan, 
Rosolowsky \& Goodman 2001). $Q(\vecr)$ is defined as the ratio of 
the rms signal within a velocity window $W$ and the rms noise, $N$
(over all velocity channels),
\begin{equation}
Q(\vecr)=\frac{1}{N}\sqrt{\frac{\sum_v T(\vecr,v)^2dv}{W}},
\label{4}
\end{equation}
where $dv$ is the width of the velocity channels.

In the present work we compute the SCF of both observational and 
synthetic spectral maps, obtained by computing the radiative transfer 
through the three--dimensional density and velocity fields of 
numerical simulations of super--sonic MHD turbulence. The result is 
typically a power law for $S_0(\dr)$ that extends up to a separation 
$\dr$ comparable to the map size, reflecting the self--similarity of 
super--sonic turbulence (Padoan, Rosolowsky \& Goodman 2001). The power 
law behavior is sometimes interrupted at an intermediate scale, possibly
suggesting the presence of a physical mechanism limiting the inertial
range of turbulence. An example of a SCF that defines an intermediate
scale is the SCF of the HI survey of the Large Magellanic Cloud (LMC)
by Kim et al. (1998, 1999). Padoan et al. (2001) have recently been able
to map the gas disk thickness of the LMC, assuming it is related to 
the intermediate scale defined by the break in the SCF power law.   

\nocite{Kim+98} \nocite{Kim+99}

\section{The Observational Data}

The absolute value of $S_0(\dr)$ at any $\dr\ $ and the slope of the
$S_0(\dr)$ power law for any given region depends on which molecular
tracer is used (Padoan, Rosolowsky \& Goodman 2001). 
Transitions probing higher gas density produce more 
fragmented integrated intensity maps than
transitions probing lower gas density, and their SCF is therefore
steeper. In order to compare the SCF of observational and synthetic 
maps it is therefore important to solve the radiative transfer
through the model density and velocity fields accurately for 
the same molecular transition that is observed. 

In this work our aim is to compute the SCF of observational data 
in order to provide constraints for theoretical models. The best
constraints come from computing the SCF of spectral maps of a 
specific molecular transition over a large range of line width 
and linear size. Observationally, small scale and narrow line 
width objects are usually mapped out with high density tracers, 
while larger objects are instead usually probed with lower density 
tracers. $^{13}$CO provides a good compromise, since it is the 
only molecule for which very large maps containing thousands of 
spectra have been obtained with a significant range of resolution. 
In this work we have therefore chosen to use observational and 
synthetic maps of the J=1--0 line of $^{13}$CO.

We have used 11 $^{13}$CO maps. For each map, we have
listed in Table~1 the approximate size, the distance, 
the rms velocity computed as the standard deviation
of the line profile averaged over the whole map, the telescope beam 
size, the spatial sampling, the width of the velocity channels
and the spectral quality defined in the previous section.
Smaller maps have been obtained from portions of the maps of the 
Taurus, Perseus and Rosette molecular cloud complexes and the SCF
has been computed for each of them. The position of these smaller 
maps within the molecular cloud complexes is shown in Figures~1 and 2. 
They have been called T1 to T7 in Taurus, P1 to P5 in Perseus, 
R1B in the Rosette Molecular cloud map by Blitz \& Stark (1986) and 
R1 and R2 in the Rosette Molecular cloud map by Heyer et al. (2001).
\footnote{More conventional designations for some of the subregions are given
in Table~2}

\nocite{Blitz+Stark86} \nocite{Mizuno+95} \nocite{Arce+Goodman2001}
\nocite{Bensch+2001}

The SCF of each map has been approximated with a power law, over the 
range of spatial separations where a power law fit is relevant.
For each power law fit we compute its slope, \alp, 
and its absolute value at 1~pc, \scf:
\begin{equation} 
S_0(\ell)=S_0(1\,pc)\,\left(\frac{\ell}{1\,pc}\right)^{-\alpha}
\label{alphadef}
\end{equation}
The values of \alp, \scf, $\sigma_v$ (the line of sight rms velocity)
and the galactic coordinates of the center of each map are given in 
Table~2. The SCF of maps of molecular cloud complexes 
and some smaller regions are shown in Figure~\ref{fig5}.

\section{MHD Simulations and Synthetic Spectral Maps}

We solve the compressible MHD equations in a staggered mesh of 
$128^3$ computational cells,
with volume centered mass density and thermal energy, face centered 
velocity and magnetic field components, edge centered electric 
currents and electric fields and with periodic boundary conditions.
The code uses shock and current sheet capturing techniques to ensure
that magnetic and viscous dissipation at the smallest resolved scales 
provide the necessary dissipation paths for magnetic and kinetic energy.
A more detailed presentation of the numerical method can be found elsewhere. 
(Padoan \& Nordlund 1999). 

For the purpose of the present work we have computed numerical solutions
of the MHD equations using an isothermal equation of state, and a random
driving force. In all experiments, the initial density is uniform, and the 
initial velocity is random. We generate the velocity field in Fourier space,
and we give power, with a normal distribution, only to the Fourier
components in the shell of wave-numbers $1 \leq k L / 2 \pi \leq 2$. 
We perform a Helmholtz decomposition, and use only the solenoidal 
component of the initial velocity. However, a compressional component
of the velocity field develops almost immediately due to the 
flow compressibility. The external driving force is generated 
on large scales in the same way as the velocity field.
The initial magnetic field is uniform, and is oriented parallel to the 
$z$ axis: ${\mathbf B}=B_0{\mathbf {\hat{z}}}$. 

Because of the limited numerical resolution we have chosen not to model
the collapse of turbulent density fluctuations. Self--gravity has therefore
been neglected. We have recently started to compute turbulent 
self--gravitating flows with a numerical mesh of 500$^3$ cells. 
Results of the analysis of these larger simulations including 
self--gravity will be presented in future works.

\subsection{Numerical Models}

We have run a number of MHD simulations
in a $128^3$ computational mesh, with periodic boundary conditions. 
The simulations are intended to describe the turbulent dynamics in 
the interior of molecular clouds. The two most important 
numerical parameters in the models are the rms sonic and Alfv\'{e}nic
Mach numbers, $M_{\rm S}$ and $M_{\rm A}$. The rms sonic Mach number is 
here defined as the ratio of the rms flow velocity and the speed of sound.
The Alfv\'{e}nic Mach number is defined as the ratio of the rms flow
velocity and the Alfv\'{e}n velocity, $v_{\rm A}=B/\sqrt{4\pi\rho}$, 
where B is the volume--averaged magnetic field strength.

All the models used in 
this work have $M_{\rm A}=10$, except for model E that has $M_{\rm A}=1$,
according to the suggestion that 
the dynamics of molecular clouds is essentially super--Alfv\'{e}nic
(Padoan \& Nordlund 1999). Our numerical simulations conserve magnetic flux,
and so the volume averaged magnetic field is constant in time. As a consequence,
also the value of $M_{\rm A}$ as defined above remains constant. 
However, the value of $B^2$ grows with time (until equilibrium is reached) 
due to compression and stretching of magnetic field lines
(see Padoan \& Nordlund 1999). If we define the Alfv\'{e}n
velocity using the rms value of the magnetic field strength, instead of its
volume average, then the typical Alfv\'{e}nic Mach number in our 
super--Alfv\'{e}nic runs is $M_{\rm A}\approx 2$, because of the formation 
of regions with large value of magnetic field strength (mainly dense regions, 
as found in observations).

The sonic Mach number of observed turbulent motions in molecular 
clouds is $M_{\rm S}> 10$ on the scale of several parsecs, and
decreases toward smaller scale. The turbulent velocity becomes 
comparable to the speed of sound only on very small scale, 
$\le 0.1$~pc. In order to study the effect of the sonic Mach
number on the SCF, we have computed MHD models with different values
of $M_{\rm S}$, $M_{\rm S}=10$, 5, 2.5, 1.25 and 0.625. Each model has been 
run for approximately six dynamical times (the dynamical time is here 
defined as the ratio of half the size of the computational box and the 
rms flow velocity), in order to achieve a statistically relaxed state, 
independent of the initial conditions. 

The velocity and density fields from the final snapshot of each model 
have been used to compute $^{13}$CO $J=1-0$ spectra, solving the radiative 
transfer with a non--LTE Monte Carlo code (\S~4.2).
While the MHD calculations are independent of the physical value of
the average gas density, the size of the computational mesh (or the column
density) and the kinetic temperature, these physical parameters are 
necessary inputs for the radiative transfer calculations. 
  
\nocite{Falgarone+92}

The models are scaled to physical units assuming a value for
i) the kinetic temperature, $T_{\rm K}$, that determines the physical unit
of velocity (the numerical unit of velocity is the speed of sound); 
ii) the average gas density, $\langle n \rangle$; iii) the size of 
the computational
box, $L_0$. For all models we have assumed $T_{\rm K}=10$~K, typical of
molecular clouds. The dependence of observed average gas density and cloud
size on the observed rms turbulent velocity (or sonic Mach number, assuming
a constant value of $T_{\rm K}$) is well--approximated by empirical Larson type
relations (Larson 1981). However, the size--velocity relation has 
a large intrinsic scatter (Falgarone, Puget \& Perault 1992), and both the 
size--velocity and density--size relations have been criticized by several authors
(Loren 1989; Kegel 1989; Scalo 1990; Issa, MacLaren \& Wolfendale 1990; 
Adler \& Roberts 1992; Vazquez-Semadeni, Ballesteros-Paredes \& Rodriguez 1997; 
Ostriker, Stone \& Gammie 2001; Ballesteros--Paredes \& Mac Low 2002). 
For these reasons, we scale the MHD models in 
four different ways. These four sets of models are all based on the same
five MHD turbulence models and differ from each other only in the way they
are rescaled to physical units when computing the radiative transfer. 
Models A1 to A5 and B1 to B5 have all the same 
value of the average density, $\langle n \rangle = 300$~cm$^{-3}$.
Models A1 to A5 have all the same size $L_0=5$~pc and column density
$N_{\rm col}=4.5\times 10^{21}$~cm$^{-2}$; models B1 to B5 have
$L_0=20$~pc and $N_{col}=1.8\times 10^{22}$~cm$^{-2}$. 
Models A1R to A5R and B1R to B5R are rescaled using the Larson
type relations:
\nocite{Larson81} \nocite{Vazquez-Semadeni+97Larson}
\begin{equation}
M_{\rm S}=M_{\rm S,1pc}\left(\frac{L}{1pc}\right)^{0.5}
\label{larson1}
\end{equation}
where a temperature $T_{\rm K}=10$~K is assumed, and
\begin{equation}
\langle n \rangle=n_{1{\rm pc}}\left(\frac{L}{1pc}\right)^{-1}
\label{larson2}
\end{equation} 
that is equivalent to a constant mean surface density.
Models A1R to A5R have the same column density as models
A1 to A5, that is $n_{1{\rm pc}}=1.5\times10^3$~cm$^{-3}$ 
in equation (\ref{larson2}); they also have sizes $L_0=$10, 
2.5, 0.625, 0.156 and 0.039 pc respectively, which implies 
$M_{\rm S,1pc}=3.16$ in equation (\ref{larson1}). 
Models B1R to B2R have the same column density as models B1 
to B5, that is $n_{1{\rm pc}}=6.0\times10^3$~cm$^{-3}$ in 
equation (\ref{larson2}); they have sizes $L_0=$20, 5, 1.25, 
0.31 and 0.078 pc respectively, which implies 
$M_{\rm S,1pc}=2.23$ in equation (\ref{larson1}). 
Finally, the equipartition model (model E) has been computed 
only for one value of the rms sonic Mach number, $M_{\rm S}=10$.
It is rescaled to the Larson type relations only once, for a size 
of 10~pc and a column density of $N_{\rm col}=4.5\times 10^{21}$~cm$^{-2}$. 
For this model we have computed spectral maps along 5 different 
directions, three orthogonal to the faces of the numerical mesh,
as in the other experiments, and two along diagonal directions,
at an angle of 54.7$^{\rm o}$ with the average magnetic field ($z$ axis).

Maps from diagonal directions sample lines of sight of different length
at different map positions (longer at the central position than near the
corners). However, the number of computational cells along each line of
sight is on the average even larger than in maps from orthogonal
directions, since a diagonal line of sight often cuts through the
computational cells away from their center (close to their corners).
Furthermore, the maps are computed only for a region of size equivalent
to that of maps from orthogonal directions ($90\times90$ cells),  
eliminating the corners of the computational mesh. As a result, only a few 
percent of the spectra are generated from lines of sight sampling less than
50 computational cells. A fraction of the the lines of sight close to
the map edges are nevertheless shorter than the energy injection scale
(approximately half the size of the computational mesh). This may
introduce a bias toward smaller line width, since velocity differences
are expected to grow with increasing distances. This bias
or its effect on the SCF should be small, since our results seem to
vary smoothly as a function of the angle between the line of sight and
the direction of the average magnetic field.

The models A4, A5, B3, B4 and B5 have velocity dispersion significantly
smaller than found observationally at the scale of 5~pc. They are not
used here to test the validity of models with such low velocity
dispersion, but rather to test the ability of the SCF method to
rule them out as poor description of molecular cloud turbulence.

In order to test the ability of the SCF to rule out unphysical
models, we have also computed two stochastic models, S2 and S4.
In both models the density field is a random field with a Log--Normal
probability distribution function, and a power law power spectrum
with power law exponent equal to -1 (the approximate value found in
our MHD models). The velocity field is generated
as a Gaussian field, also with power law power spectrum. The power law
exponent of the velocity field power spectrum is -2 (close to the actual
value in the MHD models) in model S2 and -4 in model S4. For the purpose
of computing the radiative transfer and the synthetic spectral maps, both 
models have been scaled to a physical size $L_0=20$~pc and a column density
$N_{\rm col}=4.5\times10^{21}$~cm$^{-2}$.

These two stochastic models are unphysical in the sense that they
are not solutions of the fluid equations. Statistical properties 
such as the power spectrum and the probability density function of 
density and velocity may be similar to those of flows obtained by 
solving the fluid equations, but their phase correlations are unphysical. 
This is in part illustrated by the fact that these stochastic models look
clumpy, rather than filamentary as real clouds and MHD models. Furthermore,
their velocity and density fields cannot be self--consistent because they
are computed independently of each other. It is shown below that the SCF
method can indeed rule out these unphysical models. 

The sonic rms Mach number, $M_{\rm S}$, the average gas density, 
$\langle n \rangle$, and the physical size, $L_0$, 
of the different models used for the radiative transfer computations 
are given in Table~3.

\subsection{Radiative Transfer and Synthetic Spectral Maps}

The radiative transfer calculations were carried out with a Monte Carlo
program which is a generalization of the one-dimensional Monte Carlo method
(Bernes 1979) into three dimensions. The model cloud is divided into small,
cubic cells in which physical properties are assumed to be constant. The
discretization allows the inclusion of arbitrary kinetic temperature and
molecular abundance variations. However, in the present calculations, the
temperature and relative abundances are kept constant. The 2.73~K cosmic
background is used as the external radiation field. There are important
differences between our program and the normal Monte Carlo method, and some
principles of the implementation are given below. A detailed description is
given elsewhere (Juvela 1997).

\nocite{Bernes79} \nocite{Juvela97} 

In the basic Monte Carlo method radiation field is simulated with photon
packages, each representing a number of real photons. The packages are created
at random velocities at random locations and sent toward random directions.
Each package is followed through the cloud and interactions between photons
and molecules are counted. Later this information is used to solve new
estimates for the level populations of the molecules.

In our method the radiative transfer is simulated along random lines going
through the cloud. Initially, as a photon package enters the cloud it contains
only background photons. As the package goes through a cell in the cloud some
photons emitted by this cell are added to the package and, in particular, the
number of photons absorbed within the emitting cell is calculated explicitly.
This becomes important when cells are optically thick and, compared with
normal Monte Carlo simulation, ensures more accurate estimation of the energy
transfer between cells. In our program each simulated photon package
represents intensity of all simulated transitions and Doppler shifts at the
same time. The number of individual photon packages is correspondingly
smaller, and in the present case we use 240\,000 photon packages per
iteration. The lines are divided into 70 fixed velocity channels. There is no
noise associated with random sampling of Doppler shifts. The simulated
velocity range was adjusted according to the velocity range found in the model
clouds. The channels are narrow compared with the total line widths and 
smaller than
or equal to the smallest intrinsic line widths in the cells. The velocity
discretization is therefore not expected to affect the results of the
calculations.

The density and velocity fields from the MHD simulations are sampled on
a numerical mesh of $128^3$ cells. To speed up the radiative transfer calculations
the density and velocity fields were rebinned into a mesh of $90^3$ cells
by linear interpolation. The velocity dispersion between neighboring cells
in the original $128^3$ data cube was used to approximate the turbulent
line width within each cell of the new $90^3$ data cube. This
velocity dispersion should apply to a scale slightly larger than the size
of the cells in the $90^3$ mesh. However, this is approximately compensated
by the fact that numerical dissipation in the MHD simulations decreases 
significantly the velocity dispersion on very small scale, below the actual 
turbulent inertial--range value at that scale.

On each iteration new level populations are solved from the equilibrium
equations and iterations are stopped when the relative change is below
$\sim$2.0$\cdot$10$^{-4}$ in all cells. Only the six lowest levels were tested for
convergence. The relative changes tend to be largest on the upper levels where
the level populations become very small and, on the average, the convergence
of the relevant first energy levels is much better than the quoted limit. The
total number of energy levels included in the calculations was nine, a number
clearly sufficient in case of excitation temperatures below 10\,K. The
collisional coefficients were taken from Flower \& Launay (1985) and Green \&
Thaddeus (1976).

\nocite{Flower+Launay85} \nocite{Green+Thaddeus76}

The final level populations were used to calculate maps of 90$\times$90
spectra toward three directions perpendicular to the faces of the MHD data
cube. For the equipartition model E spectra were calculated also along 
two diagonal directions. In
these cases the maps of 90$\times$90 spectra do not extend over the whole
projected cloud area. Each spectrum corresponds to the intensity calculated
along one line of sight (spectra are not convolved with a larger beam). The
spectra contain 60 velocity channels as in the Monte Carlo simulation.
The results were compared with spectra calculated assuming LTE conditions. The
comparison showed that for typical physical conditions found in molecular
clouds the LTE assumption would be unsuitable (Padoan et al. 2000).

\nocite{Padoan+99per}

\section{Observational Correlations and Theoretical Models}

Most observational and theoretical spectral maps yield a SCF
that can be approximated by a single power law within a range
of spatial separations, often spanning over an order of magnitude.
From each power law fit we compute its slope, \alp, and its
value at 1~pc, \scf, defined as in (\ref{alphadef}). 
We also compute the value of the velocity 
dispersion, $\sigma_v$, from each map, measured as the standard 
deviation of the $^{13}$CO $J=1-0$ spectrum averaged over the 
entire map. The values of \alp, \scf\ and $\sigma_v$ computed 
from the observational maps are given in Table~2, while the values 
of the same quantities from the theoretical models are given in Table~3. 

Every model provides three sets of values, because spectral maps 
have been computed using three orthogonal directions for the line 
of sight. Each group of three sets of values can be interpreted 
as the same model cloud being ``observed'' from different directions,
or as three different model clouds with comparable rms velocity.
In the equipartition model, E, the rms velocities
inferred from different directions are very different from each other,
the largest rms velocity being found in the direction parallel to the
mean magnetic field (along the $z$ axis), and the lowest in the directions
perpendicular to the magnetic field. For this model we have also computed
spectral maps from two more lines of sight, corresponding to diagonal
directions across the computational box.   

Figure~\ref{fig6} shows the SCF of the model A1R, with line of sight
parallel to the direction of the mean magnetic field. The SCF of the 
equipartition model E is also shown for four lines of sight, 
two diagonal, one parallel to the direction of the mean magnetic field
and one perpendicular to it to it. The figure shows that the SCF of model 
E is very
sensitive to the line of sight, due to the large variations of the rms
velocity in different directions relative to the mean magnetic field.

The velocity dispersion relative to the speed of sound, or the value
of the sonic Mach number, is the most important physical 
parameter characterizing the nature of the turbulence. We
therefore study the dependence of the SCF on the
turbulent velocity dispersion (or the rms sonic Mach number, \mach)
and propose to use this dependence to test theoretical models against
the observational data.

In Figure~\ref{fig7} the slope of the SCF is plotted against the line of
sight velocity dispersion. 
The top panels show the models of constant size and constant 
average gas density (models A1 to A5 -left, and B1 to B5 -right); the bottom
panels show the models rescaled according to Larson type relations
(models A1R to A5R -left and B1R to B5R -right). The observations indicate
a strong correlation between \alp\ and $\sigma_v$, over an order of magnitude
in $\sigma_v$. A least square fit to the observational data gives:
\begin{equation}
\alpha_{\rm obs}=0.30\,\sigma_{v,{\rm obs}}^{0.37 \pm 0.07},
\label{empirical1}
\end{equation}
and for the super--sonic and super--Alfv\'{e}nic 
models rescaled with Larson type relations (A1R to A5R):
\begin{equation}
\alpha_{{\rm MHD}}=0.31\,\sigma_{v,{\rm MHD}}^{0.47 \pm 0.04},
\label{models1}
\end{equation}
consistent with the observational result (the uncertainty in the
exponent is the standard deviation from the least square fit).
The corresponding models not scaled with the Larson type relations
(A1 to A5) are also indistinguishable from the observational result 
(see Figure~\ref{fig7} top left panel). Models of type B 
(right panels of Figure~\ref{fig7}) have instead values of \alp\
significantly smaller than the average ones from the observational
data. The stochastic
models S2 and S4 are indistinguishable from each other; they are also
totally inconsistent with the empirical \alp--$\sigma_v$ relation,
which allows them to be ruled out as invalid by the SCF method.
Finally, the equipartition model E provides values that are consistent
with the observations, and comparable to the super--Alfv\'{e}nic models,
apart from a larger scatter of values between different lines of sight. 

We interpret the increase of \alp\ with $\sigma_v$ as a consequence of the
increasing compressibility of the turbulent flow ($\sigma_v$ is roughly
proportional to the rms sonic Mach number of the flow because the
temperature in all the models is $T_{\rm K}=10$~K, and approximately the same
in the observed regions). The value of \alp\ is in general found to grow with
increasing density contrast, probably due to the increasing concentration
of the mass along the line of sight around one or few dense cores, which
helps decorrelating the spectra from each other. 
    
The value of \scf\ is plotted against $\sigma_v$ in Figure~\ref{fig8}.
The top panels show the models of constant size and the bottom panels
the models scaled with the Larson type relations, as in Figure~\ref{fig7}.
The values of \scf\ and $\sigma_v$ from the observational maps are weakly
correlated, with \scf\ slightly increasing with increasing $\sigma_v$:
\begin{equation}
S_{0,{\rm obs}}({\rm 1pc})=0.45\,\sigma_{v,{\rm obs}}^{0.13\pm 0.08}
\label{empirical2}
\end{equation}
A tight correlation is instead found in the models of constant size
(top panels of Figure~\ref{fig8}), with \scf\ decreasing with increasing 
$\sigma_v$. This inconsistency between the models and the observations
is most likely due to the fact that molecular clouds of 5 to 20 pc of size
(as assumed by these models) are never found with velocity dispersion 
as low as assumed in models A4, A5 and B3, B4 and B5.  
The bottom panels of
Figure~\ref{fig8} show that the inconsistency is in fact mostly resolved 
as soon as the model sizes are scaled according to the Larson type
relation. For the models A1R to A5R we obtain:
\begin{equation}
S_{0,{\rm MHD}}({\rm 1pc})=0.39\,\sigma_{v,\rm{MHD}}^{-0.03\pm 0.05}
\label{MHD2}
\end{equation}
If models with $\sigma_v<0.2$~km/s were not included (justified by the
absence of such low velocity dispersions in the observational sample),
the slope of the least square fit would be $0.07\pm 0.04$, fully
consistent with the observations.
The equipartition model yields values of \scf\ and $\sigma_v$ consistent
with the observations in all directions, but the one parallel to the 
mean magnetic field. It could be concluded that either none of the 
observed objects has a significant component of the magnetic field
along the line of sight, or that all of them have a magnetic field
weaker than predicted by the equipartition model, consistent
with the super--Alfv\'{e}nic models.

In Figure~\ref{fig9} we have plotted observations and models 
on the \alp--\scf\ plane.
The constant size models are again inconsistent with the observations,
as is expected since the observational maps span a large range 
of scales. When the models are scaled according with the Larson
type relations and the realistic average column density of 
$4.5\times 10^{21}$~cm$^{-2}$ (Myers \& Goodman 1988), the observed scatter in the 
\alp--\scf\ plane is reproduced. The trend of the absolute value of
\alp\ to increase with \scf\, for large values of both of them is
also reproduced, between models with rms Mach 5 and 10 (A2R and A1R
respectively); however, models with rms Mach of 20 or 30 would be 
necessary to fit the \alp\ and \scf\ values measured for the Rosette
molecular cloud, which can be appropriately resolved only with a 
numerical resolution in excess of $256^3$ computational cells.

\nocite{Myers+Goodman88Larson}

While the stochastic models S2 and S4 are only marginally inconsistent
with the observations in this plot, the line of sight parallel to the direction
of the mean magnetic field and one of the two diagonal lines of sight in the equipartition model E are again
inconsistent with the observational data.

\section{Discussion}

The SCF has been proposed as a statistical tool to test the validity
of theoretical models describing the structure and dynamics of star 
forming clouds (Rosolowsky et al. 1999). 
In a previous work we improved the SCF method 
by studying its dependence on spatial and velocity resolution and on 
instrumental noise (Padoan, Rosolowsky \& Goodman 2001). Here we have 
applied that improved SCF to a number of large $^{13}$CO maps 
of molecular cloud complexes and obtained empirical correlations 
that can be used to test theoretical models. Of the theoretical 
models we have computed some compare well with the empirical 
correlations and some do not, which shows that the SCF can be used
as an effective tool to rule out inappropriate or unphysical models. 

The empirical correlations we have obtained relate the values of \alp ,
\scf\ and $\sigma_v$ with each other. The \alp--$\sigma_v$ correlation rules 
out the unphysical stochastic models (S2 and S4). Such models were found to 
produce spectral line profiles similar to observational ones by  
Dubinski, Narayan \& Phillips (1995). They have also been used 
as models of the density field in molecular clouds by Stutzki 
et al. (1998) and to calibrate their principal component analysis
by Brunt \& Heyer (2002). The SCF \alp--$\sigma_v$ correlation 
shows that stochastic models are inappropriate to describe the 
structure of molecular cloud complexes. 

\nocite{Dubinski+95} \nocite{Stutzki+98} \nocite{Brunt+Heyer2002I}

Models not scaled with Larson type relations (A1--A5, B1--B5)
and models with larger--than--average column density (B1--B5, B1R--B5R) 
have also been compared with the empirical SCF correlations to
show that incorrectly scaled models are readily ruled out by the 
SCF method. 

The \scf--$\sigma_v$ and the \alp--\scf\ correlations do not favor 
the model with equipartition of kinetic and magnetic energies
(model E). Such model yields too small values of \scf\ or too large
values of \alp\ compared with
the observational data, when seen in the direction parallel to the
average magnetic field. Of the two diagonal directions, one is consistent
with the data and the other is not. 

A possible interpretation is that 
none of the observed regions has an average magnetic field oriented 
close to the direction of the line of sight. The equipartition 
model starts to be inconsistent with the observational data when seen
along the diagonal directions, at an angle of 54.7$^{\rm o}$ to the average 
magnetic field. The line of sight should be within such an angle to the
magnetic field in approximately 40\% of the cases, assuming random 
orientation of the average magnetic field in the observed regions.

An alternative interpretation is that all the 
observed regions have an average magnetic field strength smaller
than in the equipartition model, and consistent with super--Alfv\'{e}nic
conditions. The super--Alfv\'{e}nic models rescaled with Larson type 
relations are in fact able to reproduce the empirical SCF correlations. 
However, the total number of truly independent directions on the sky in 
the present observational sample is still small. More regions should 
be studied to rule out the equipartition model based on the SCF results.

The analysis of the MHD models could in principle
give different results if self--gravity was taken into account. 
However, the introduction of self--gravity is not expected to
decrease the value of \alp\ and increase the value of \scf ,
as necessary to make the equipartition model consistent with
the observational \alp--\scf\ correlation. The main effect of 
self--gravity is the collapse of the densest regions, increasing the 
density contrast beyond the level due to the turbulence alone. 
This could slightly increase the value of \alp\ because
we interpret the increase of \alp\ with $\sigma_v$ in the MHD models as due to the 
increased compressibility of the turbulent flow. An increase in the value 
of \scf\ is not expected because the local collapse of dense cores 
cannot increase the correlation between spectra at large distances (for a given
value of \alp\, an increase in \scf\ would correspond to an increase 
in the SCF at large spatial separation). Nevertheless, the effect of 
self--gravity should be tested by including it in the numerical solution 
of the MHD equations. The numerical resolution should also be larger
than in the present work to resolve the initial phase of the gravitational
collapse of dense cores. We have only recently started to compute
self--gravitating flows in a numerical mesh of $500^3$ cells, and their
analysis will be presented in future works. 

Padoan \& Nordlund (1997, 1999) have proposed that the dynamics 
of molecular clouds on large scales is consistent with super--Alfv\'{e}nic 
turbulence and inconsistent with the equipartition model. In numerical
simulations of super-Alfv\'{e}nic turbulence the 
average magnetic energy grows with time, even if flux is conserved
(the average magnetic field is constant). The magnetic field
strength is increased locally mainly in regions of compression in
super--sonic turbulence, and in part by stretching of field lines.
Even if initial conditions are such that the turbulence is highly
super-Alfv\'{e}nic, magnetic pressure is often larger than 
thermal pressure in the postshock gas, due to the amplification
of the magnetic field components perpendicular to the shock direction.   
Equipartition of dynamic pressure, $\rho\, v^2$, and 
magnetic pressure, $B^2/8\pi$, is therefore achieved locally, but
not necessarily over the whole flow. For example, in super--sonic and 
super-Alfv\'{e}nic runs at a resolution of $250^3$, the ratio of volume--average 
magnetic and dynamic pressures relaxes at a value 
$\langle P_{\rm m}\rangle/\langle P_{\rm d}\rangle\approx 0.12$,
starting from initial conditions with 
$\langle P_{\rm m}\rangle_{\rm in}/\langle P_{\rm d}\rangle_{\rm in}\approx 0.005$
(Padoan et al. 2003). Comparable values are found
in the numerical experiments used in this work.
The amplification of the magnetic field by the turbulence therefore
does not alter the super--Alfv\'{e}nic character of the flow.

The correlation between local magnetic field strength and gas density
in super--sonic and super--Alfv\'{e}nic turbulence has a very large scatter,
and a well defined upper envelope with $B\propto \rho^{\,0.4}$, both
consistent with the observational data (Padoan \& Nordlund 1997, 1999;
Ostriker, Stone \& Gammie 2001; Passot \& Vazquez--Semadeni 2002).
The largest values of the magnetic field strength are generally found
in dense cores, but some dense cores may have a relatively weak magnetic
field. However, dense cores assembled by turbulent shocks are not
expected to have internal super--Alfv\'{e}nic turbulence, because
of the dissipation of kinetic energy in the shocks and of the
amplification of the magnetic field in the compressed gas. Observational
evidence for an approximate equipartition of turbulent and magnetic energy
in dense cores would therefore not be inconsistent with the 
super--Alfv\'{e}nic character of the large scale flow that assembles them.

The comparison between our theoretical models and the observational
data could be improved if more regions with sub--sonic turbulence
were available in the observational sample. Small velocity dispersion
is found in small objects, according to Larson's velocity--size
relation, or to the power spectrum of turbulence. The spatial resolution
in single dish surveys is typically too low to sample a small object
(fraction of a parsec) with a very large spectral maps (several thousands 
of spectra). The only exceptions in the observational sample used in this
work are L1512 and L134a. These two
large maps of nearby clouds with very low velocity dispersion were obtained by 
Falgarone et al. (1998) as part of their IRAM key project, focused on
regions of relatively low column density at the edges of molecular cloud
complexes.

\nocite{Falgarone+98}

Maps of large regions with very large velocity dispersion are instead
more easily obtained from observations than in numerical simulations.
Assuming a gas kinetic temperature of the order of 10~K, a line of sight  
(one dimensional) velocity dispersion in excess of 2~km/s corresponds
to a sonic rms Mach number of the flow \mach $\ge 20$. 
In the present work we have not computed numerical flows with
\mach $>10$, since that would require a larger numerical
resolution (the density contrast grows linearly with the Alfv\'{e}nic
Mach number and therefore with the value of $M_{\rm S}$ if the
average magnetic field strength is not varied). 
For this reason the models do not reach the largest values 
of $\sigma_v$, \alp\ and \scf\ obtained from the observations (from the maps
of the Rosette molecular cloud complex). The progression of models 
toward increasing values of $M_{\rm S}$ suggests that a model with 
\mach $\ge 20$ would likely fit the observed \alp--\scf\ values found 
in the Rosette molecular cloud complex, where the observed velocity
dispersion is in excess of 2~km/s. This is illustrated in 
Figure~\ref{fig10}. The top panel of Figure~\ref{fig10} 
shows the \alp--\scf\ for the 
observational data. The shaded area shows the range of values covered
by the theoretical models A1R to A5R. The bottom panel shows the same plot
for the model A1R to A5R. Each diagonal segment connects the values for 
the three directions of each model. The values of \mach\ and $L_0$ of
the models are also given in the plot. The arrow marks the direction
of increasing \mach\ suggesting that models with \mach $\approx 20$
may fit the observations with the largest velocity dispersion.

\section{Summary and Conclusions}

In the present work we have computed the spectral correlation function 
(SCF) of spectral maps of molecular cloud complexes and regions within 
them, observed in the J=1--0 transition of $^{13}$CO. We have found 
that the SCF is a power law over approximately an order of magnitude 
in spatial separation. The power law slope of the SCF, \alp, its 
normalization, \scf, and the spectral line width averaged over 
the whole map, $\sigma_v$, have been computed for all the observational 
maps. We have obtained empirical correlations between these quantities 
and have proposed to use them to test the validity of theoretical 
models of molecular clouds. 

Theoretical models of spectral line maps have been generated by computing 
the radiative transfer through the numerical solutions (density and velocity
fields) of the magneto-hydrodynamic (MHD) equations, for turbulent flows
with different values of the rms sonic and Alfv\'{e}nic Mach numbers,
and also through stochastic density and velocity fields with different
power spectra. Super-Alfv\'{e}nic MHD models rescaled according to Larson 
type relations are in the best agreement with the empirical correlations. 
Unphysical stochastic models are instead ruled out. MHD models with equipartition 
of magnetic and kinetic energy of turbulence do not reproduce the observational data
when their average magnetic field is oriented approximately parallel to the 
line of sight. Finally, MHD models not rescaled according to Larson type 
relations are also inconsistent with the observational data.

We cannot exclude the possibility that different physical models for the dynamics of 
molecular clouds, or even unphysical models, that we have not tested 
here, would satisfy the empirical correlations found in this 
work. However, we have shown that the SCF method is able to rule out
certain unphysical or incorrectly scaled models. Reproducing these 
SCF results should be considered as a necessary
(but not sufficient) condition for the validity of theoretical
models describing the structure and the dynamics of molecular clouds. 
Models for which the SCF or similar statistical tests cannot be computed to 
allow a quantitative comparison with observed spectral maps cannot be
legitimately evaluated.

The comparison between theory and observations 
presented in this work requires significant computational resources. 
Numerical simulations of three dimensional turbulent flows must be run at
large resolution and the radiative transfer has to be computed in three dimensions
in order to generate synthetic spectral maps of the observed molecular transitions. 
The type of models and the physical parameters investigated in this
work are therefore limited to a few significant cases. Future work should investigate
the SCF of a larger variety of models, including different magnetic field intensities
and flows with gravitationally collapsing cores.

\acknowledgements

We are grateful to Eve Ostriker and Jim Stone for helpful comments
on our model--data comparison. The referee report by Enrique
Vazquez--Semadeni has also contributed to improve this work.
This work was supported by an NSF Galactic Astronomy grant to
AG. The work of PP was partially performed while
PP held a National Research Council Associateship Award at the
Jet Propulsion Laboratory, California Institute of Technology.
MJ acknowledges the support of the Academy of Finland Grant no. 1011055.

\clearpage


\clearpage

\onecolumn

{\bf Figure captions:} \\

{\bf Table \ref{T1}} Main parameters of the observed spectral maps: 
Approximate size, distance, rms velocity over the whole map, 
telescope beam, spatial sampling, velocity channel width, average 
spectrum quality and bibliographic reference.  \\

{\bf Table \ref{T2}} Spectral line width averaged over the whole map, 
$\sigma_v$, power law slope of the SCF, \alp\ and SCF normalization, 
\scf, galactic longitude, $l$, and galactic latitude, $b$, of the center
of all the observed maps and selected regions within them.\\

{\bf Table \ref{T3}} First three columns from the left: Model name,
rms sonic Mach number of the flow and physical size of the computational
mesh. Following columns: Line of sight velocity dispersion, SCF slope
and SCF normalization, repeated for the three orthogonal directions
for which synthetic spectral maps have been computed in each model. 
Values for the diagonal directions of model E are not given (they are
within the ranges of values covered by the other three directions
parallel and perpendicular to the mean magnetic field). \\

\begin{figure}
\centerline{\epsfxsize=13cm \epsfbox{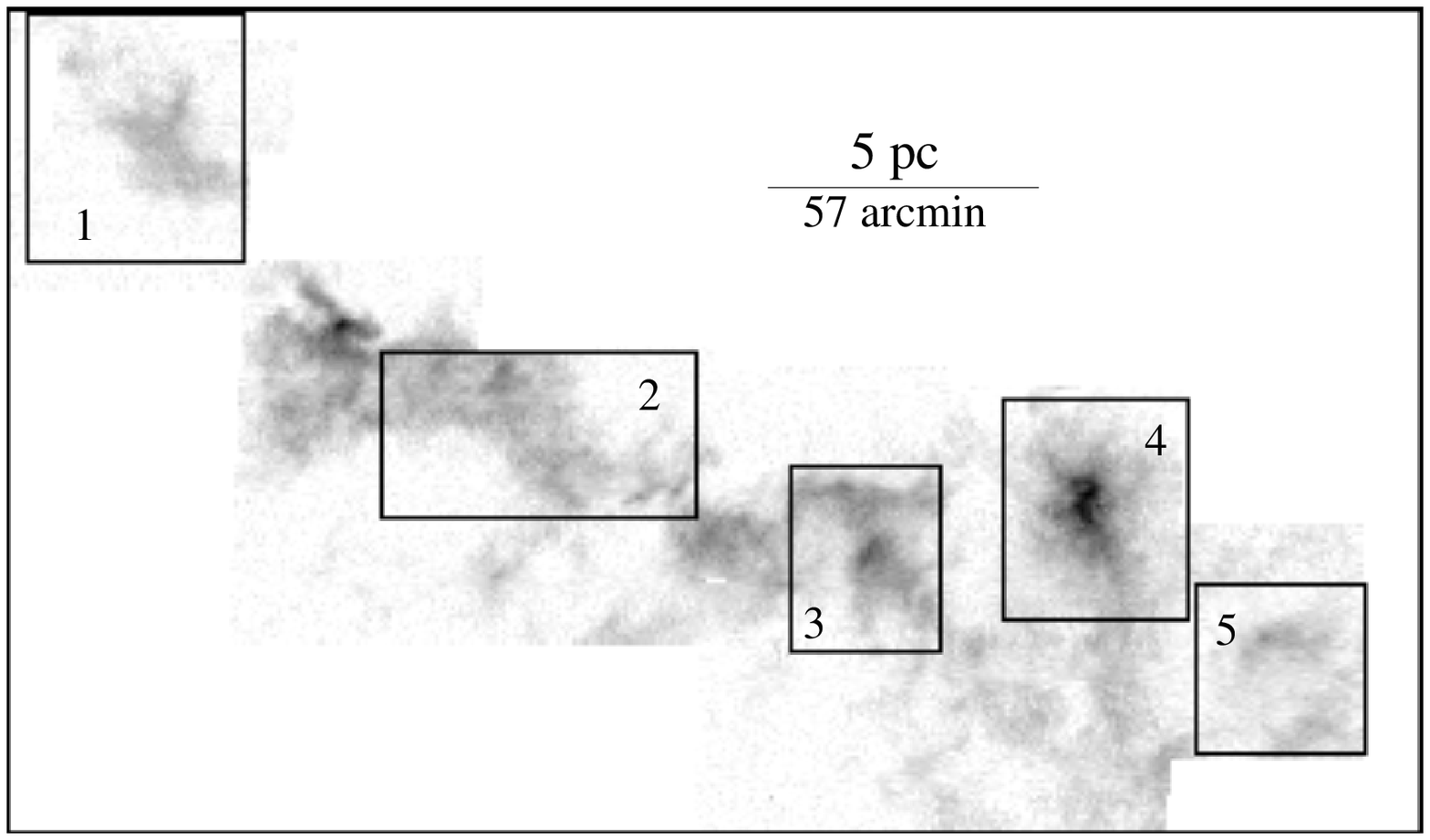}}
\centerline{\epsfxsize=13cm \epsfbox{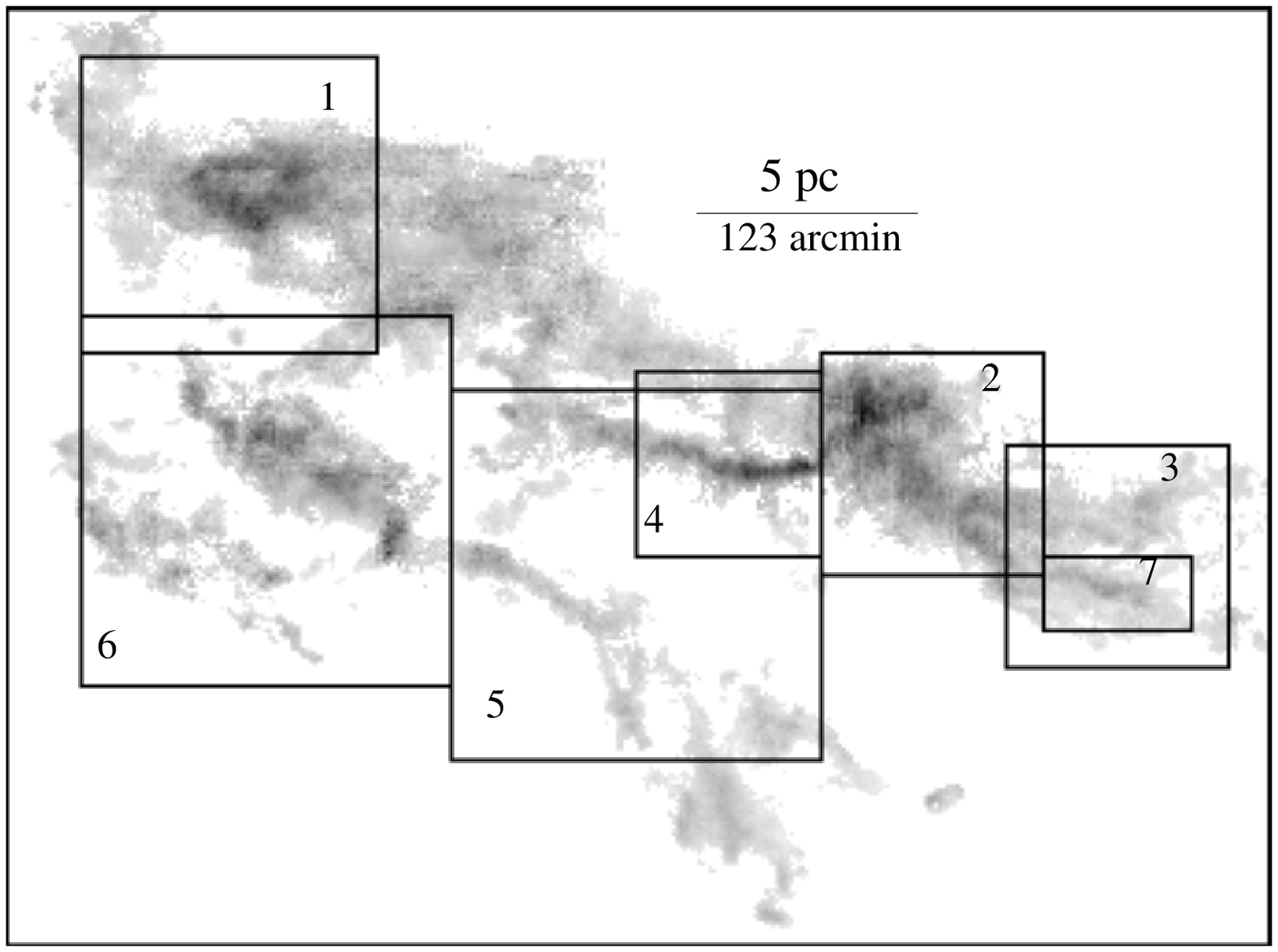}}
\caption[]{}
\label{fig1}
\end{figure}

{\bf Figure \ref{fig1}:} Top panel: Velocity integrated intensity map of 
the Perseus molecular cloud complex in the J=1-0 transition of $^{13}$CO 
(Padoan et al. 1999). Bottom panel: Same as top panel, but for
the Taurus molecular cloud complex (Mizuno et al. 1995). Smaller 
regions within the maps where the SCF has also been computed are
highlighted.\\

\begin{figure}
\centerline{\epsfxsize=13cm \epsfbox{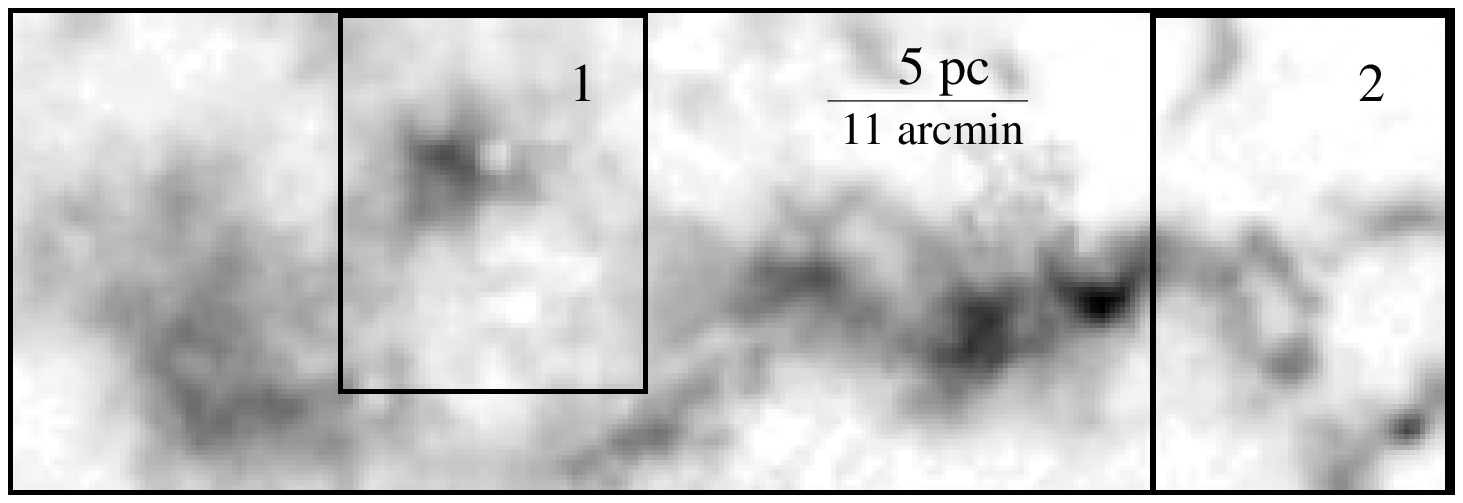}}
\vspace{1cm}
\centerline{\epsfxsize=12cm \epsfbox{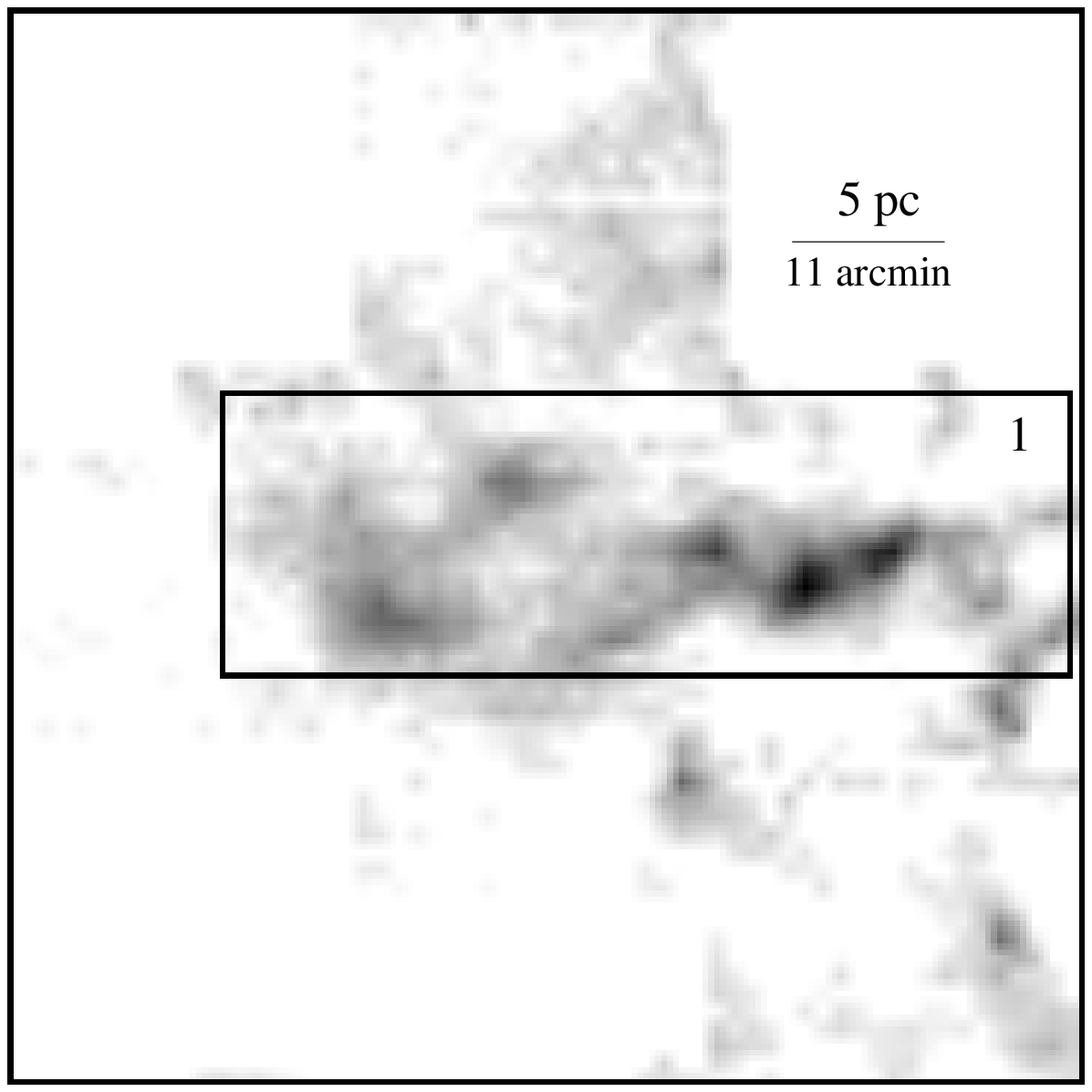}}
\caption[]{}
\label{fig2}
\end{figure}

{\bf Figure \ref{fig2}:} Same as in Figure~\ref{fig1}, but for the 
Rosette molecular cloud complex. Top panel from Heyer et al. (2001);
bottom panel from Blitz \& Stark (1986). \\

\begin{figure}
\centerline{\epsfxsize=9cm \epsfbox{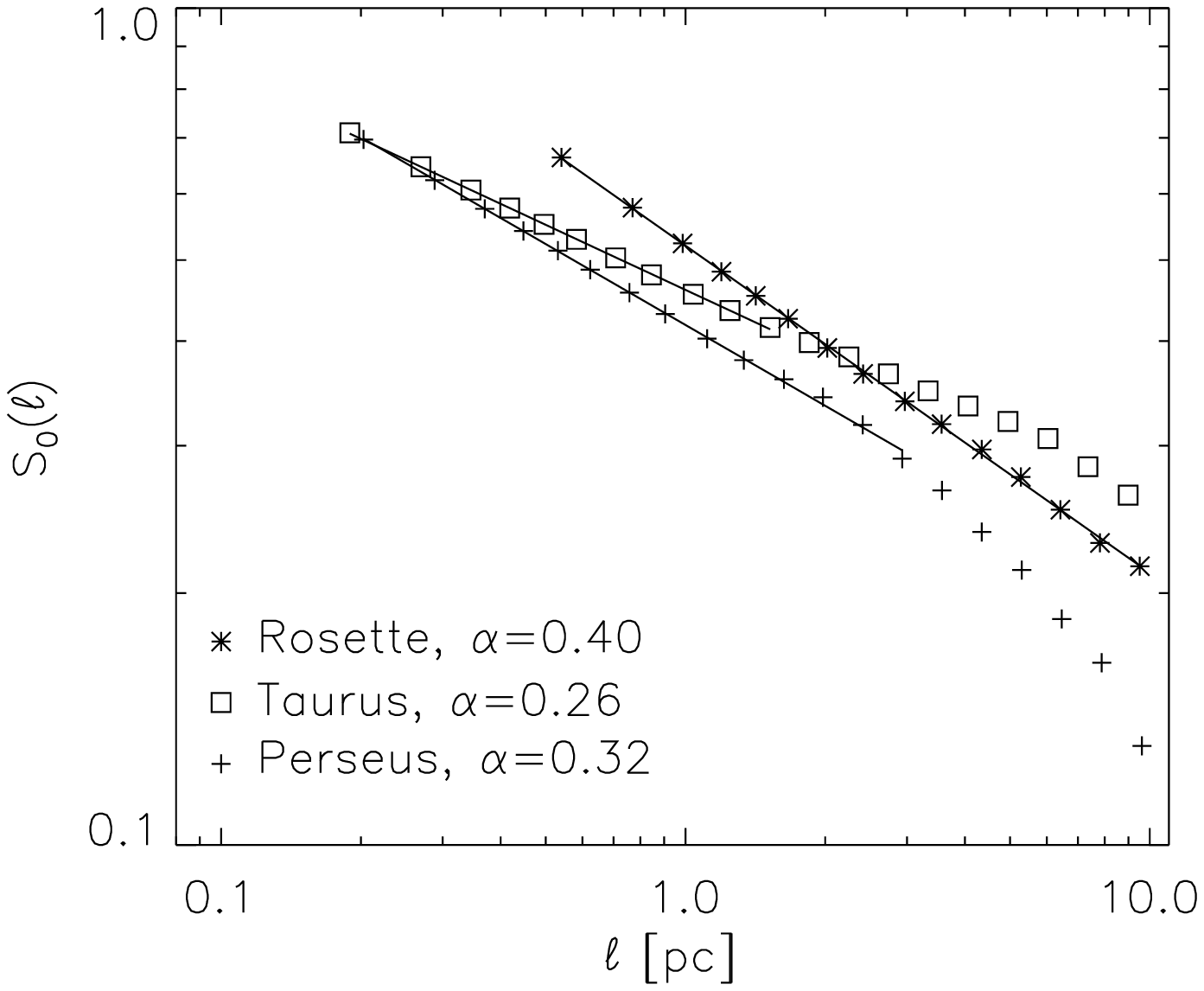}
            \epsfxsize=9cm \epsfbox{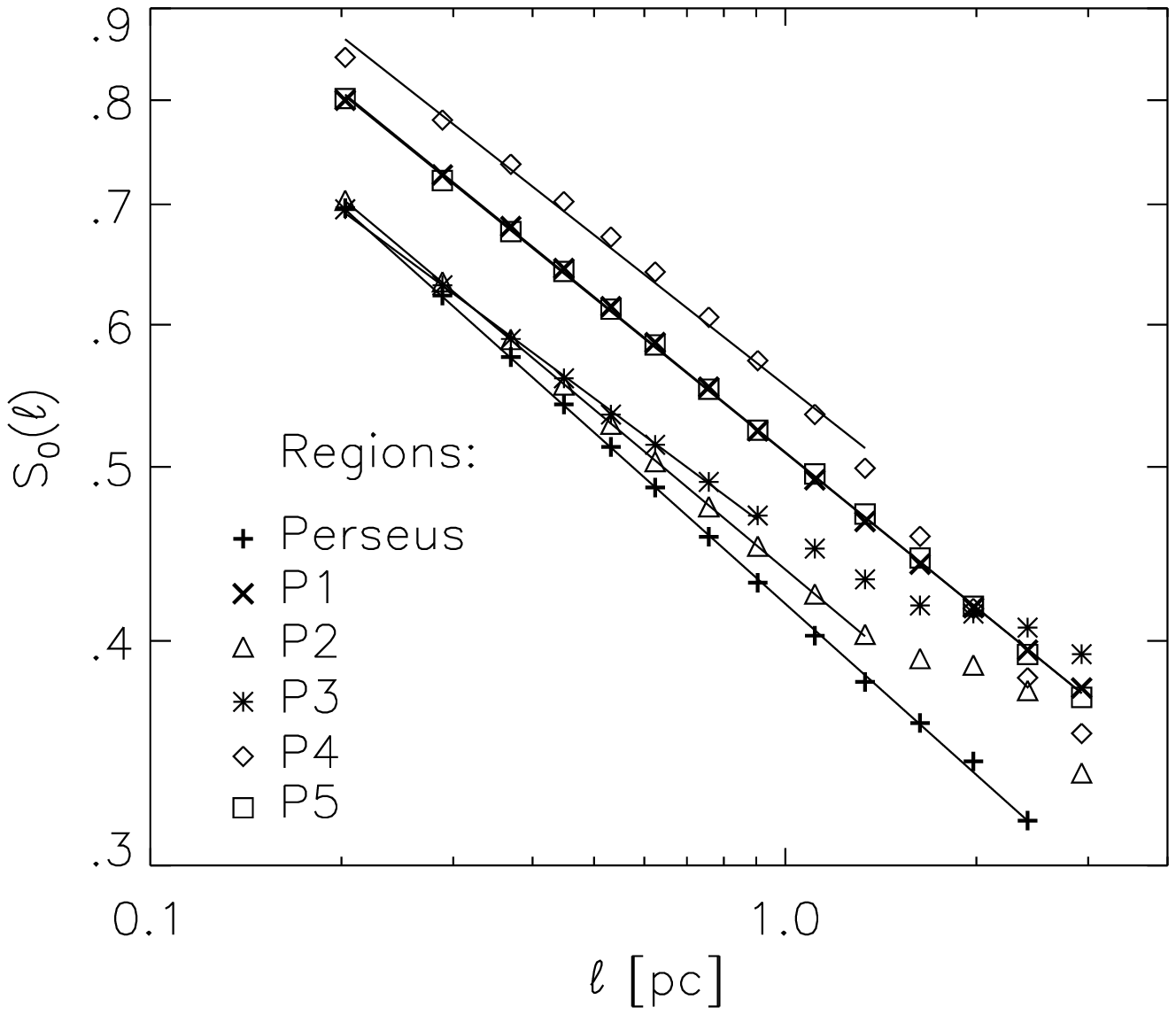}
}
\centerline{\epsfxsize=9cm \epsfbox{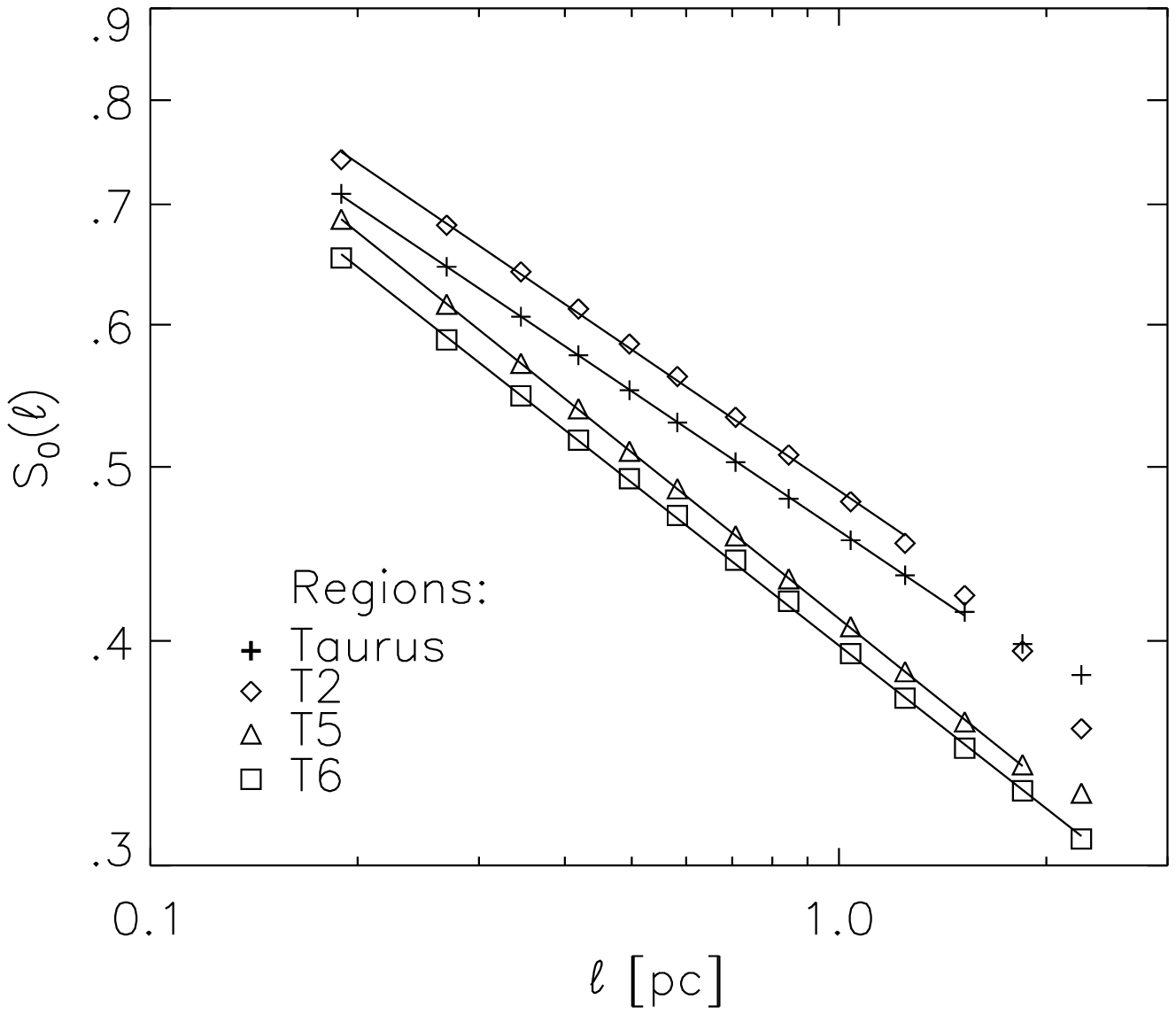}
            \epsfxsize=9cm \epsfbox{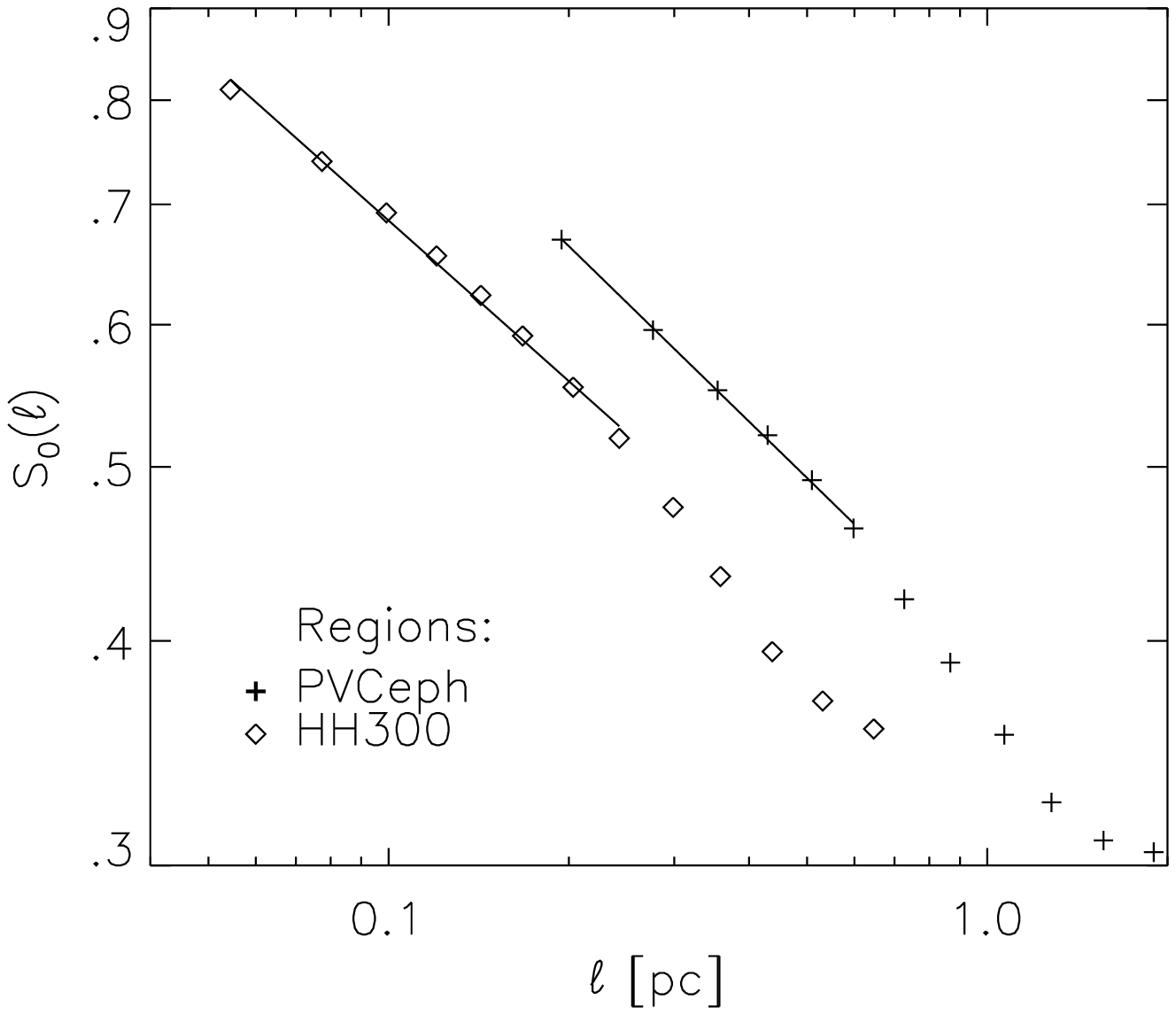}
}
\caption[]{}
\label{fig5}
\end{figure}

{\bf Figure \ref{fig5}:} Top left panel: The SCF averaged over the entire 
map of the Perseus, Rosette and Taurus molecular cloud complexes. 
Solid lines are least square fits to the power law sections of the SCF. 
The exponents $\alpha$ of the power law fits are also given in the figure. 
Top right panel: SCF of the whole map of the Perseus molecular cloud complex
and of smaller regions within the same map. Bottom left panel: SCF of the
Taurus molecular cloud complex and of smaller regions within the same map.
Bottom right panel: SCF of PVCeph and HH300. \\

\begin{figure}
\centerline{\epsfxsize=14cm \epsfbox{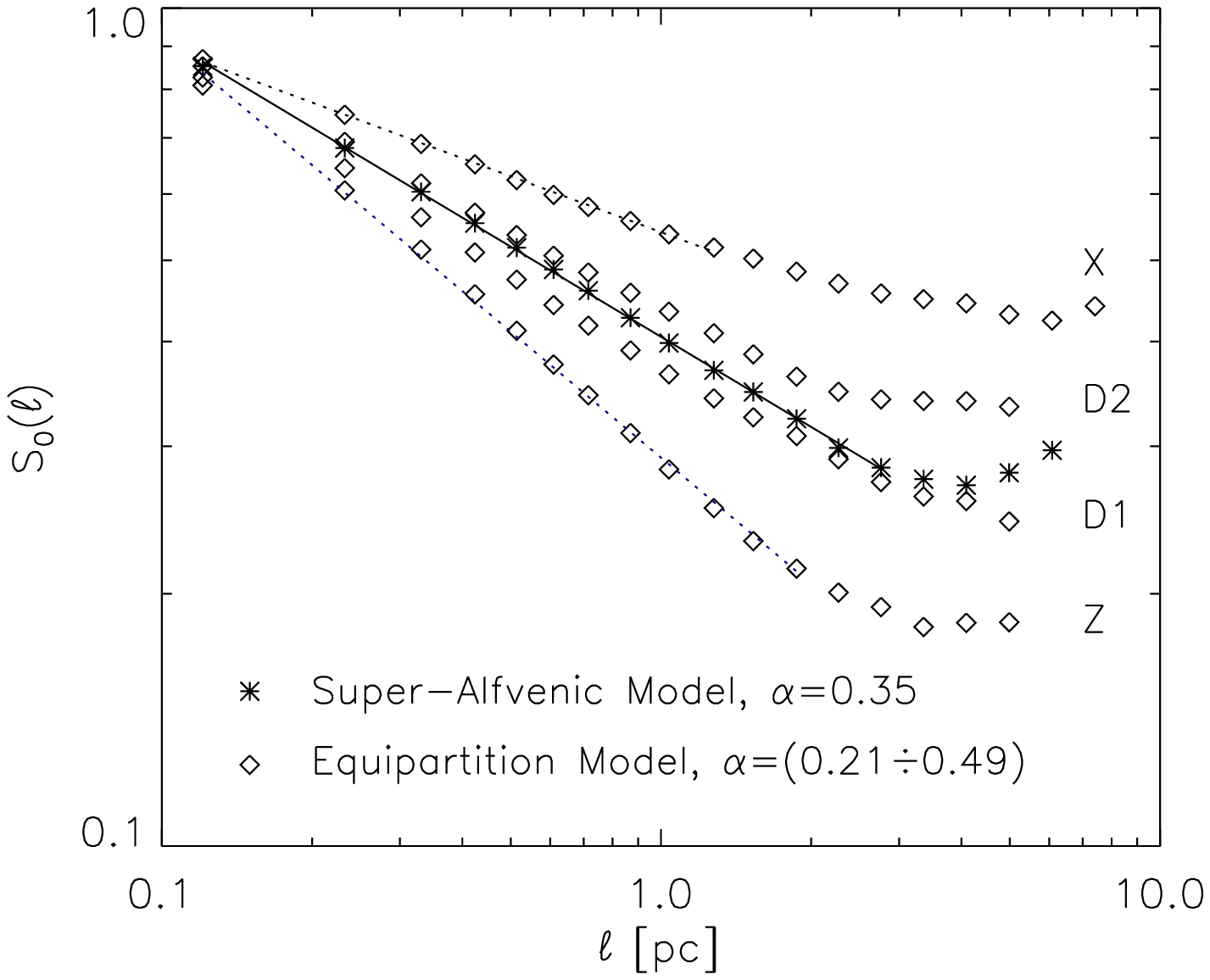}}
\caption[]{}
\label{fig6}
\end{figure}

{\bf Figure \ref{fig6}:} The SCF computed from MHD models. Asterisk symbols
are for the super-Alfv\'{e}nic model A1R in the $Z$ direction (parallel to
the mean magnetic field). Diamond symbols are for the equipartition
model in the $X$ and $Z$ direction (perpendicular and parallel to the average
magnetic field direction respectively) and along two diagonal directions ($D1$
and $D2$). The slope of the SCF increases with increasing rms velocity. The 
SCF is therefore weakly dependent on the direction of the line of sight for
the super-Alfv\'{e}nic model, while it is much steeper in the direction
parallel to the magnetic field (larger rms velocity) than in the perpendicular
direction in the equipartition model. \\

\begin{figure}
\centerline{\epsfxsize=19cm \epsfbox{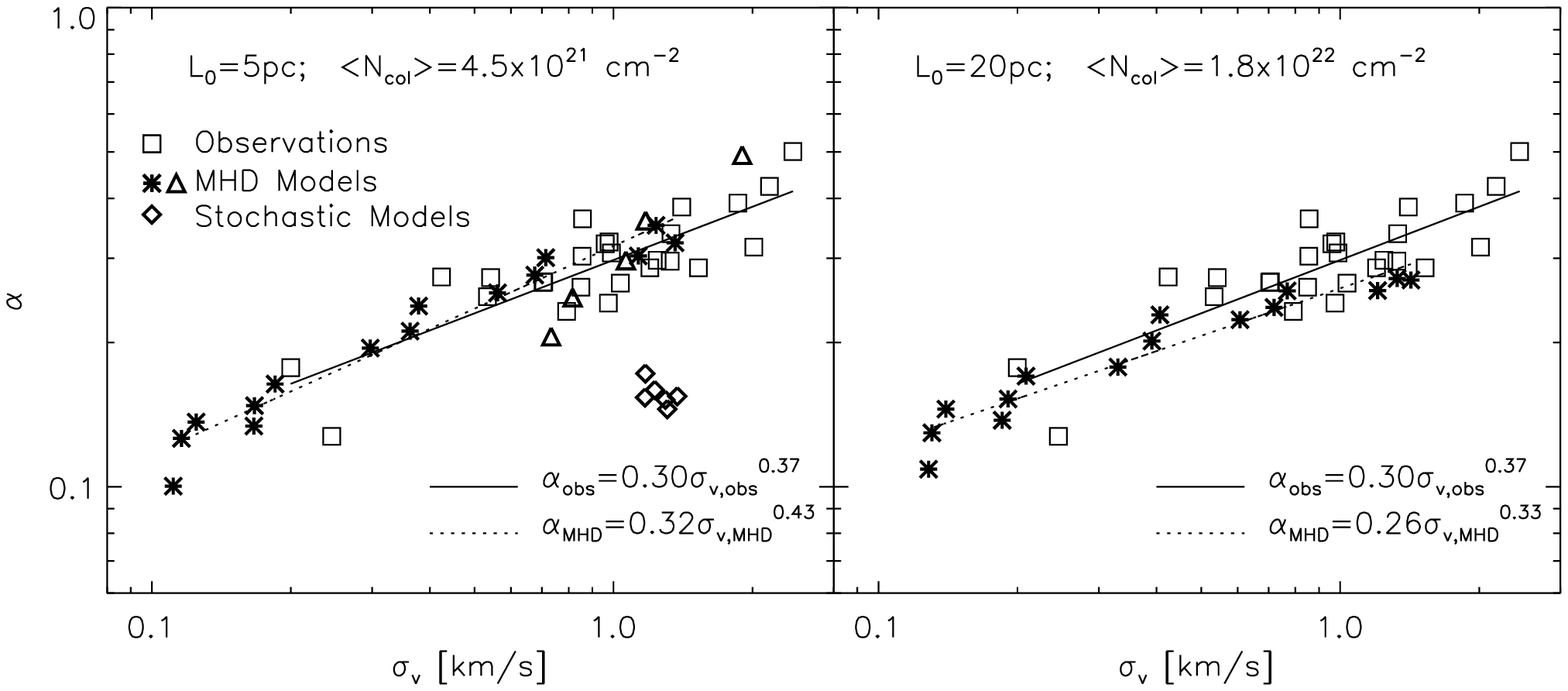}}
\centerline{\epsfxsize=19cm \epsfbox{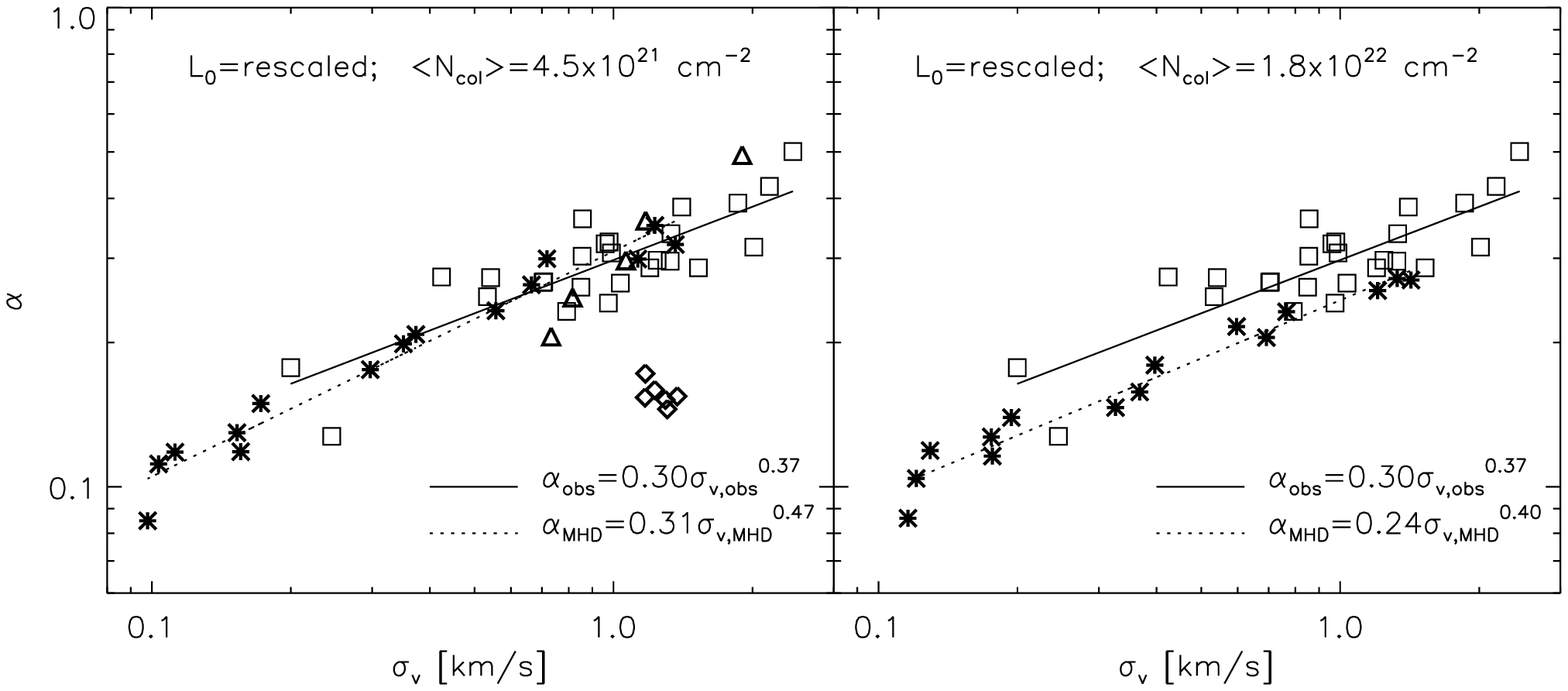}}
\caption[]{}
\label{fig7}
\end{figure}

{\bf Figure \ref{fig7}:} SCF slope versus velocity dispersion. 
The top panels show the models of constant size and constant 
average gas density as asterisks (models A1 to A5 --left, B1 
to B5 --right); the bottom panels show the models rescaled 
according to Larson type relations as asterisks (models A1R to A5R 
--left, B1R to B5R --right). Observational values are shown as 
squares, the equipartition model as triangles and the
stochastic models as diamonds. \\

\begin{figure}
\centerline{\epsfxsize=19cm \epsfbox{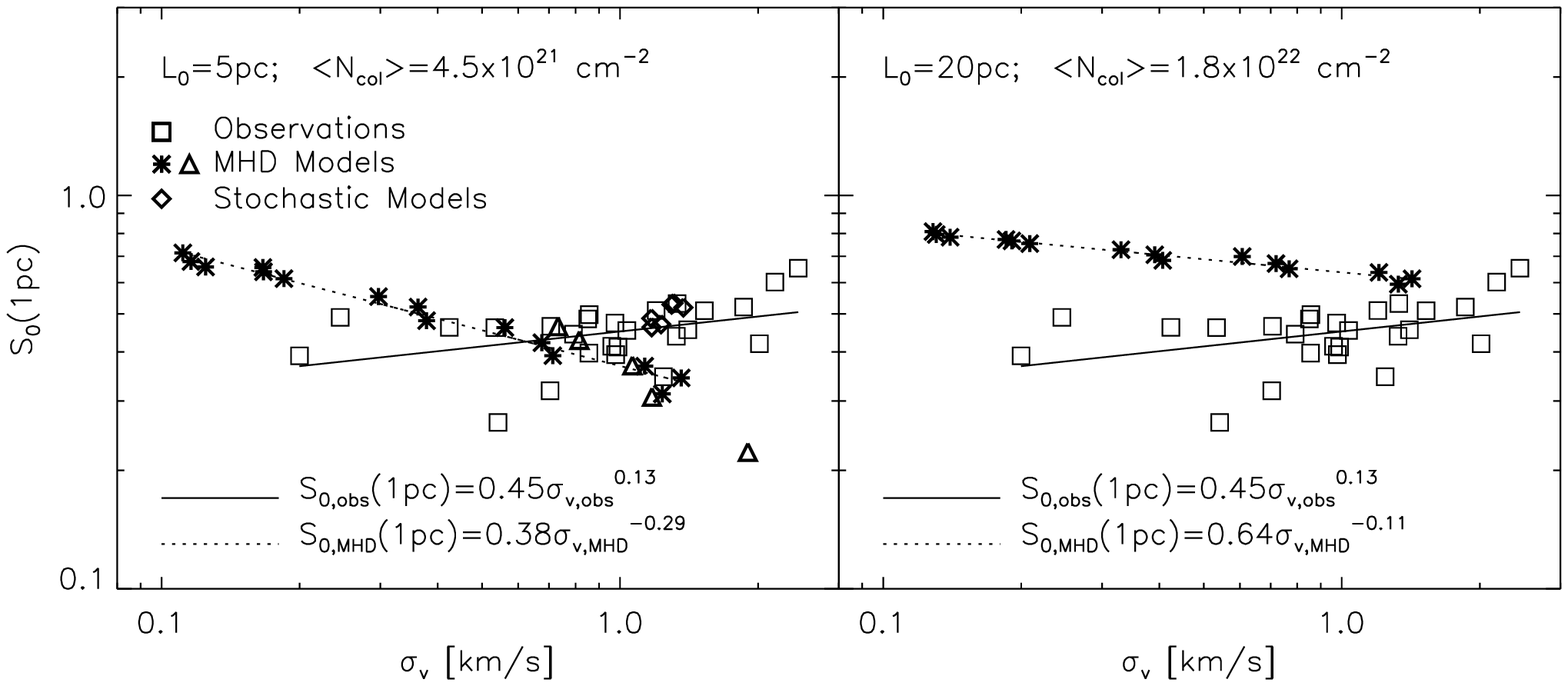}}
\centerline{\epsfxsize=19cm \epsfbox{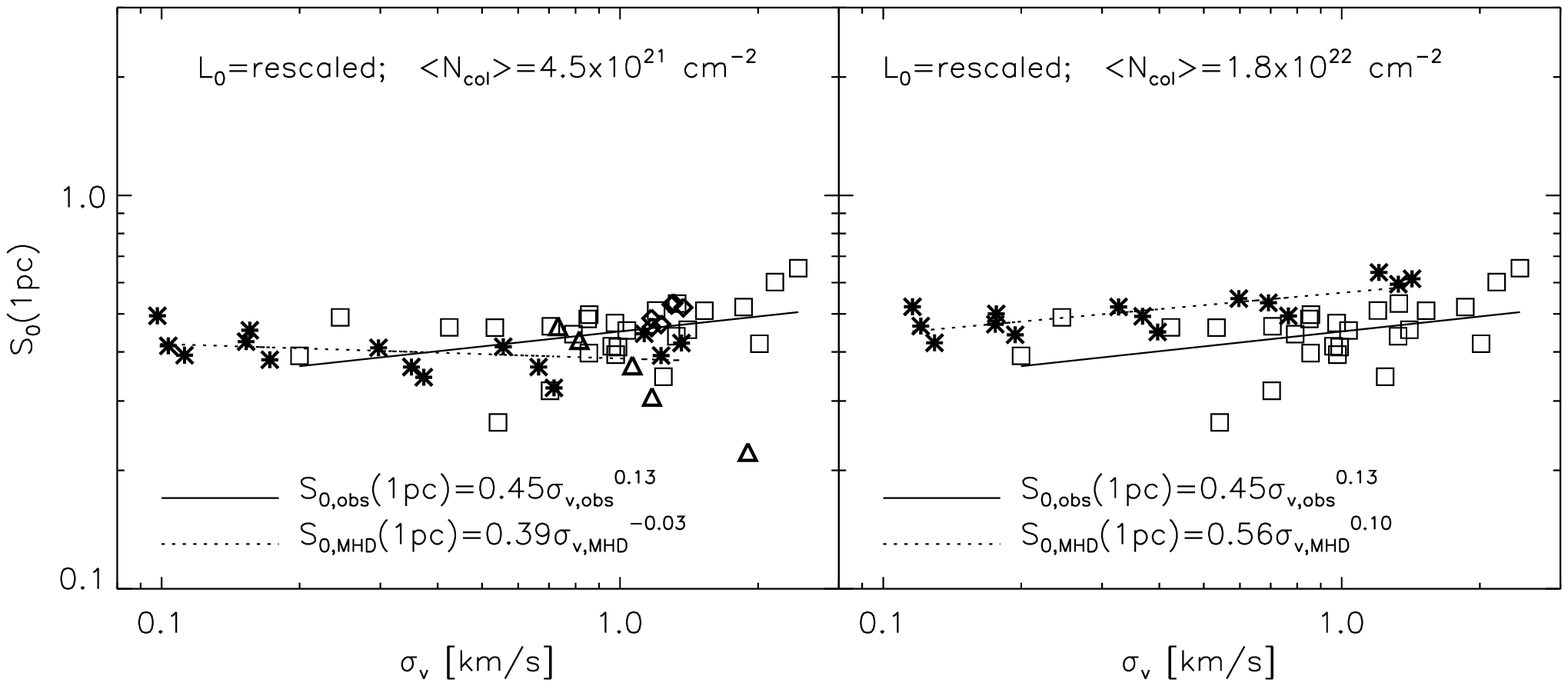}}
\caption[]{}
\label{fig8}
\end{figure}

{\bf Figure \ref{fig8}:}  SCF value at 1~pc versus velocity dispersion.
Different panels show different models as in Figure~\ref{fig7}. 
Symbols are also as in Figure~\ref{fig7}. \\

\begin{figure}
\centerline{\epsfxsize=19cm \epsfbox{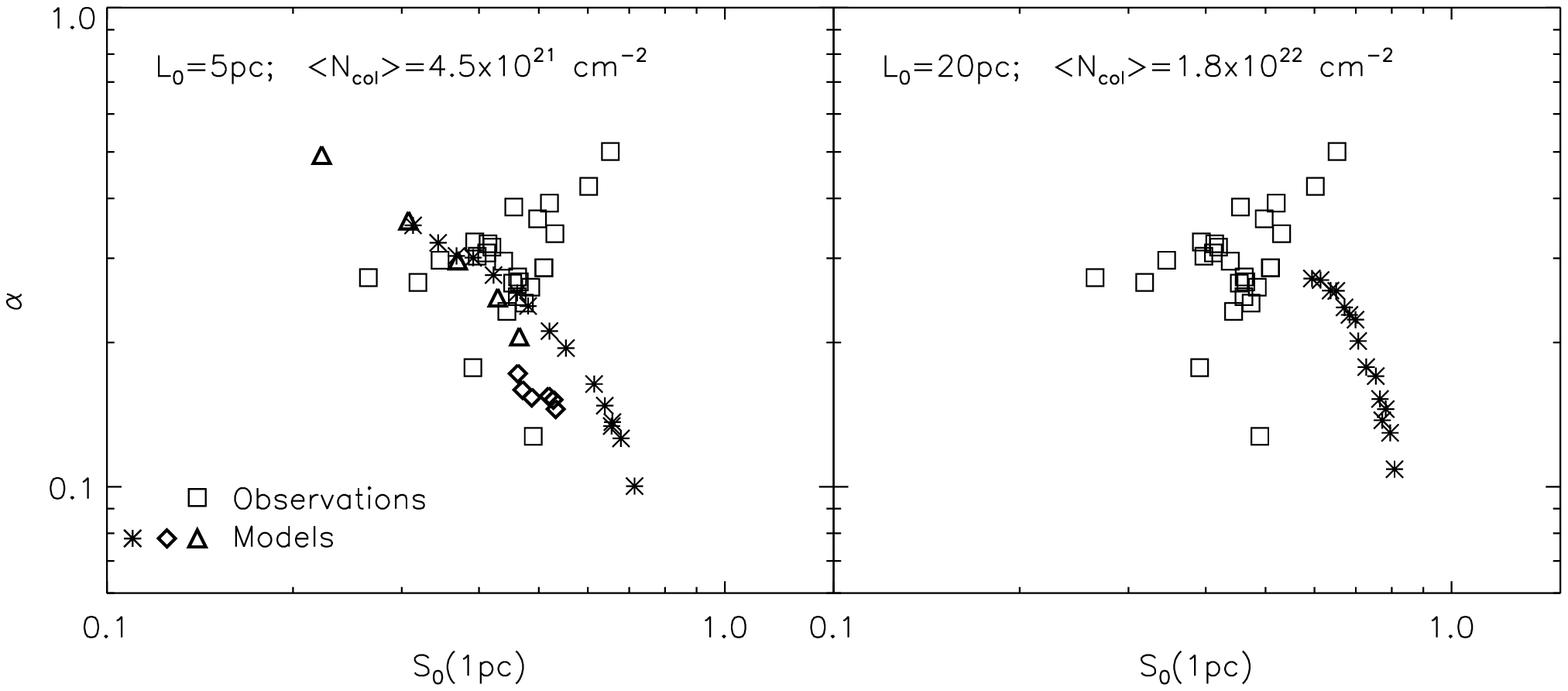}}
\centerline{\epsfxsize=19cm \epsfbox{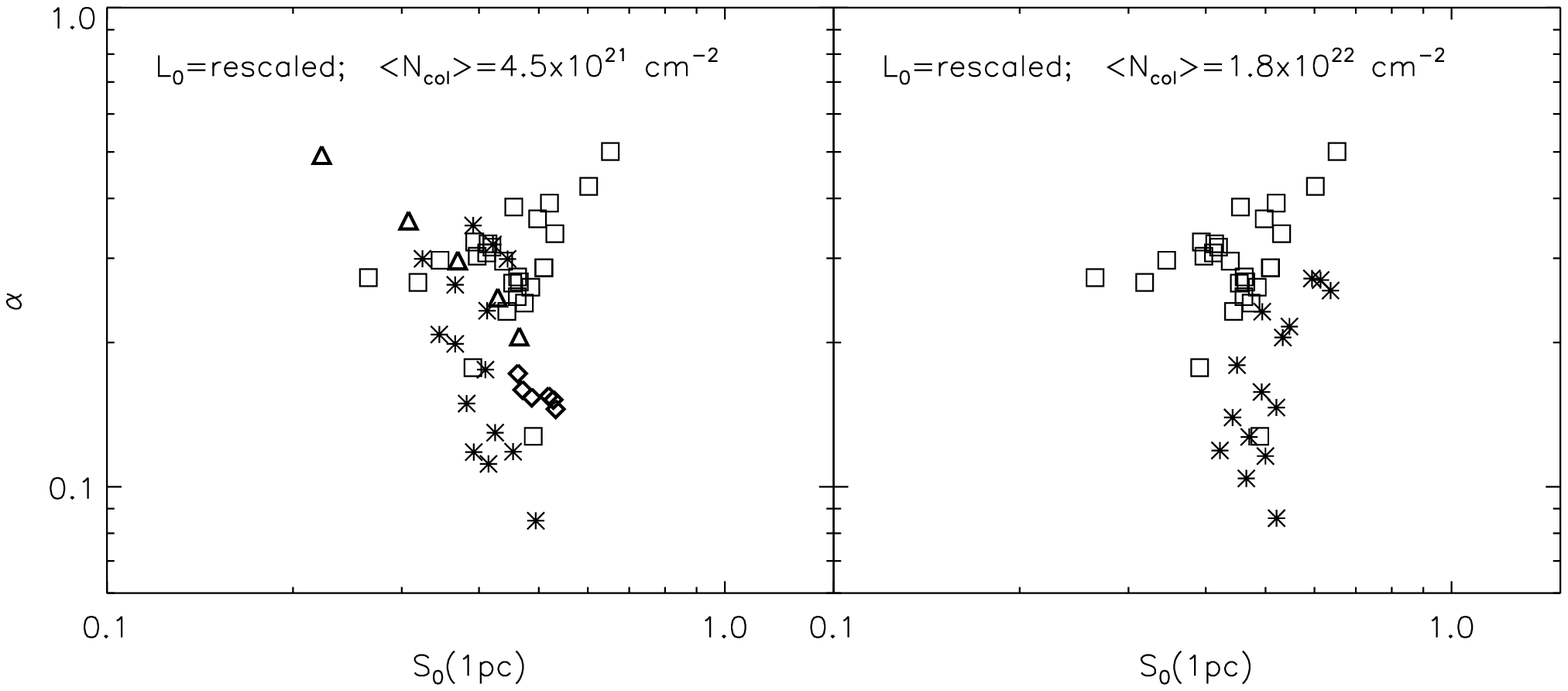}}
\caption[]{}
\label{fig9}
\end{figure}

{\bf Figure \ref{fig9}:} SCF slope versus SCF value at 1~pc. Symbols
and panels as in Figure~\ref{fig7}. \\

\begin{figure}
\centerline{\epsfxsize=15cm \epsfbox{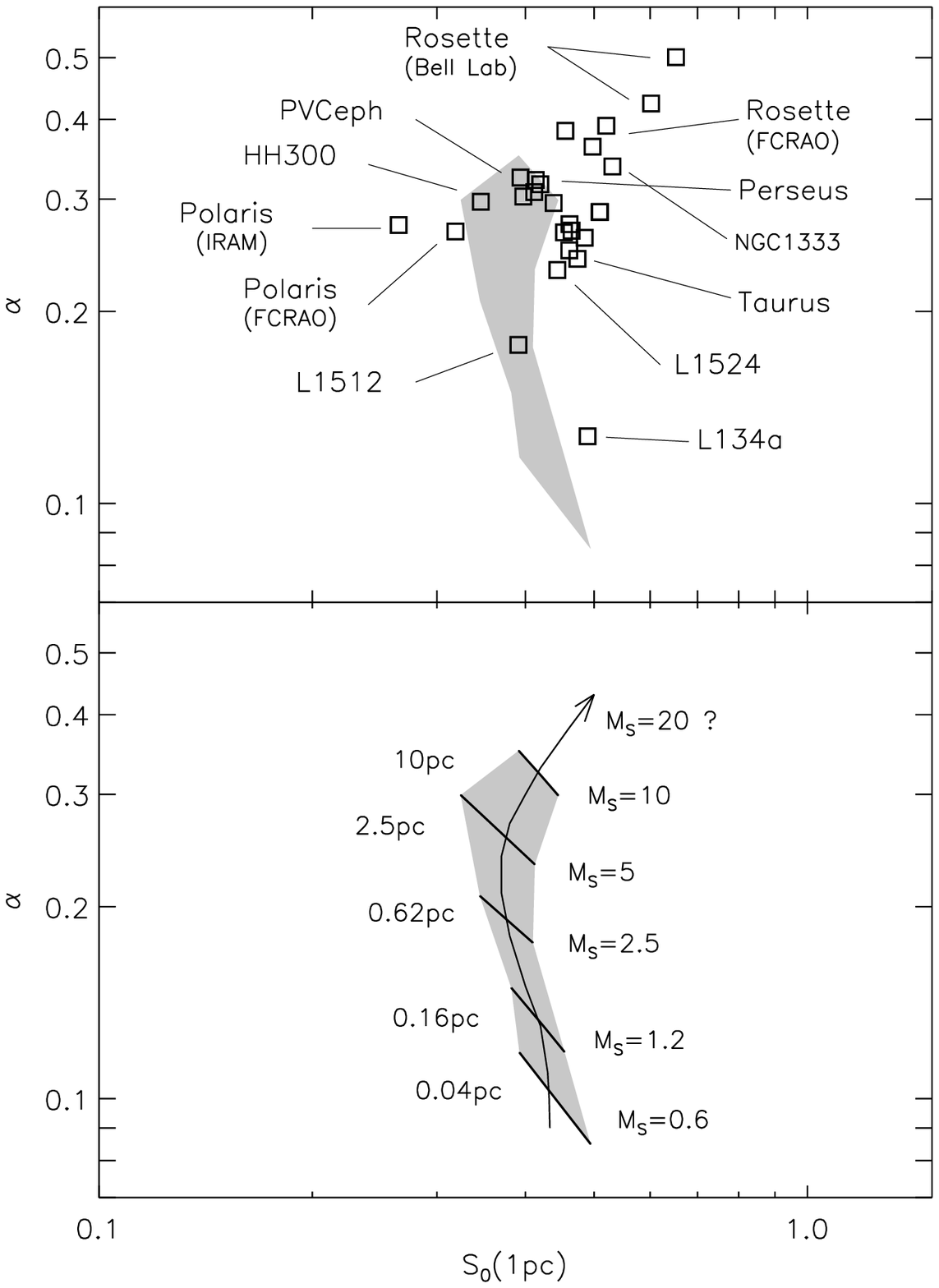}}
\caption[]{}
\label{fig10}
\end{figure}

{\bf Figure \ref{fig10}:} Top panel: Values of \alp\ and \scf\
from the observations. Some of the symbols are labeled with the region name.
The shaded area shows the range of values covered by the models A1R 
to A5R. Bottom panel: Same shaded area as in the top panel. Diagonal
segments shows the range of values of \alp\ and \scf\ for the three
directions of each model. The rms sonic Mach number of the corresponding 
model is given on the right hand side of each segment, while the value
of the linear size is given on the left hand side. The arrow indicates 
the progression of models toward larger values of sonic Mach number, \mach .


\clearpage
\begin{table}
\begin{tabular}{lcccccccl}
\hline
\hline
MC       & $L$ [pc] & D [Kpc] & $\sigma_v$ [km/s] & Beam [pc] & $dx$ [pc] & $dv$ [km/s] & $\aq$ & reference  \\ 
\hline
Taurus   &   30     & 0.14    &   0.97            &  0.11     & 0.081     &   0.10      &  2.3  & Mizuno et al. 1995   \\ 
Perseus  &   30     & 0.30    &   2.01            &  0.16     & 0.087     &   0.27      &  2.8  & Padoan et al. 1999 \\ 
Rosette  &   45     & 1.60    &   2.45            &  0.84     & 0.70      &   0.68      &  2.1  & Blitz \& Stark 1986 \\ 
Rosette  &   35     & 1.60    &   1.86            &  0.36     & 0.23      &   0.06      &  3.8  & Heyer et al. 2001 \\
L1524    &   1.5    & 0.14    &   0.79            &  0.032    & 0.015     &   0.10      &  5.2  & Bensch 2002   \\
Polaris  &   1.5    & 0.11    &   0.70            &  0.025    & 0.012     &   0.10      &  1.3  & Bensch et al. 2001   \\ 
HH300    &    2     & 0.14    &   1.24            &  0.032    & 0.023     &   0.022     &  1.5  & Arce \& Goodman 2001 \\
PVCeph   &    3     & 0.50    &   0.98            &  0.114    & 0.083     &   0.022     &  1.8  & Arce \& Goodman 2001 \\
Polaris  &   0.3    & 0.11    &   0.53            &  0.012    & 0.004     &   0.052     &  6.3  & Falgarone et al. 1998 \\
L1512    &   0.3    & 0.15    &   0.20            &  0.016    & 0.005     &   0.052     & 11.7  & Falgarone et al. 1998 \\
L134a    &   0.3    & 0.15    &   0.24            &  0.016    & 0.005     &   0.052     & 13.3  & Falgarone et al. 1998 \\  
\hline
\end{tabular}
\caption{}
\label{T1}
\end{table}

\clearpage
\begin{table}
\begin{tabular}{lcccccc}
\hline
\hline
Region         & $\sigma_v$ [km/s] & $\alpha$ & $S_0(1pc)$ & $l$  & $b$   \\ 
\hline
Taurus         &   0.97            &  0.24   &  0.47      & 170.8 & -16.2\\
T1             &   0.71            &  0.27   &  0.46      & 174.5 & -13.8\\
T2             &   0.85            &  0.26   &  0.48      & 168.0 & -16.1\\
T3             &   0.53            &  0.25   &  0.46      & 166.3 & -16.8\\ 
T4             &   0.96            &  0.32   &  0.41      & 169.8 & -16.1\\
T5             &   0.99            &  0.31   &  0.41      & 170.8 & -17.0\\
T6             &   0.85            &  0.30   &  0.40      & 174.2 & -16.3\\
T7             &   0.42            &  0.27   &  0.46      & 166.2 & -17.3\\
Perseus        &   2.01            &  0.32   &  0.42      & 160.0 & -19.3\\
P1 (B5)        &   1.20            &  0.29   &  0.51      & 161.2 & -16.8\\ 
P2             &   1.33            &  0.30   &  0.44      & 160.7 & -18.8\\  
P3 (B1)        &   1.03            &  0.27   &  0.45      & 159.8 & -20.1\\ 
P4 (NGC1333)   &   1.33            &  0.34   &  0.53      & 158.8 & -20.5\\ 
P5 (L1448)     &   1.53            &  0.29   &  0.51      & 158.6 & -21.6\\ 
Rosette (Bell Lab) &   2.45            &  0.50   &  0.65      & 207.5 & -1.8 \\
R1B            &   2.18            &  0.42   &  0.60      & 207.3 & -1.8 \\ 
Rosette (FCRAO)&   1.86            &  0.39   &  0.52      & 207.3 & -1.8 \\
R1             &   1.40            &  0.38   &  0.45      & 207.5 & -1.8 \\
R2             &   0.86            &  0.36   &  0.50      & 206.9 & -1.8 \\
L1524          &   0.79            &  0.23   &  0.44      & 173.3 & -16.3\\
Polaris (FCRAO)&   0.70            &  0.27   &  0.32      & 123.4 &  24.9\\
HH300          &   1.24            &  0.30   &  0.35      & 172.9 & -16.7\\
PVCeph         &   0.98            &  0.32   &  0.39      & 102.9 &  15.2\\
Polaris (IRAM) &   0.54            &  0.27   &  0.26      & 123.7 &  24.9\\
L1512          &   0.20            &  0.18   &  0.39      & 171.8 & -5.2 \\
L134a          &   0.24            &  0.13   &  0.49      &   4.3 & 35.8 \\  
\hline
\end{tabular}
\caption{}
\label{T2}
\end{table}

\clearpage
\begin{table}
\begin{tabular}{lcc|ccc|ccc|ccc}
\hline
\hline
      &       &       &      &   x   &      &      &   y   &      &      &   z   &        \\
\hline
Model & \mach & \L0   & \sig & \alp  & \scf & \sig & \alp  & \scf & \sig & \alp  & \scf   \\ 
\hline
\multicolumn{12}{c}{MHD Models \,\,\,( $\langle n \rangle=300$~cm$^{-3}$ )} \\
\hline
A1    & 10.0  &  5    & 1.13 & 0.30 & 0.37 & 1.36 & 0.32 & 0.34 & 1.24 & 0.35 & 0.31   \\
A2    &  5.0  &  5    & 0.56 & 0.25 & 0.46 & 0.71 & 0.30 & 0.39 & 0.67 & 0.28 & 0.42   \\
A3    &  2.5  &  5    & 0.30 & 0.19 & 0.55 & 0.38 & 0.24 & 0.48 & 0.36 & 0.21 & 0.52   \\
A4    &  1.2  &  5    & 0.17 & 0.13 & 0.66 & 0.17 & 0.15 & 0.64 & 0.18 & 0.16 & 0.61   \\
A5    &  0.6  &  5    & 0.12 & 0.13 & 0.68 & 0.11 & 0.10 & 0.71 & 0.12 & 0.14 & 0.66   \\
\hline
B1    & 10.0  &  20   & 1.21 & 0.26 & 0.64 & 1.42 & 0.27 & 0.61 & 1.33 & 0.27 & 0.59   \\ 
B2    &  5.0  &  20   & 0.61 & 0.22 & 0.70 & 0.77 & 0.26 & 0.65 & 0.72 & 0.24 & 0.67   \\ 
B3    &  2.5  &  20   & 0.33 & 0.18 & 0.73 & 0.41 & 0.23 & 0.68 & 0.39 & 0.20 & 0.71   \\ 
B4    &  1.2  &  20   & 0.18 & 0.14 & 0.77 & 0.19 & 0.15 & 0.77 & 0.21 & 0.17 & 0.75   \\
B5    &  0.6  &  20   & 0.13 & 0.13 & 0.79 & 0.13 & 0.11 & 0.81 & 0.14 & 0.14 & 0.78   \\
\hline
\multicolumn{12}{c}{Rescaled MHD Models \,\,\, ( $\langle n \rangle=6\times10^3$~cm$^{-3}\,L_{0,pc}^{-1}$ )}\\
\hline
A1R   & 10.0  &  10   & 1.13 & 0.30 & 0.44 & 1.36 & 0.32 & 0.42 & 1.23 & 0.35 & 0.39   \\
A2R   &  5.0  &  2.5  & 0.56 & 0.23 & 0.41 & 0.72 & 0.30 & 0.32 & 0.66 & 0.26 & 0.37   \\
A3R   &  2.5  &  0.62 & 0.30 & 0.18 & 0.41 & 0.37 & 0.21 & 0.34 & 0.35 & 0.20 & 0.37   \\
A4R   &  1.2  &  0.16 & 0.16 & 0.12 & 0.45 & 0.15 & 0.13 & 0.42 & 0.17 & 0.15 & 0.38   \\
A5R   &  0.6  &  0.04 & 0.10 & 0.11 & 0.41 & 0.10 & 0.08 & 0.49 & 0.11 & 0.12 & 0.39   \\
\hline
B1R   & 10.0  &  20   & 1.21 & 0.26 & 0.64 & 1.42 & 0.27 & 0.61 & 1.33 & 0.27 & 0.59   \\ 
B2R   &  5.0  &   5   & 0.60 & 0.22 & 0.55 & 0.76 & 0.23 & 0.49 & 0.69 & 0.20 & 0.53   \\
B3R   &  2.5  &  1.25 & 0.33 & 0.15 & 0.52 & 0.40 & 0.18 & 0.45 & 0.37 & 0.16 & 0.49   \\
B4R   &  1.2  &  0.31 & 0.18 & 0.12 & 0.50 & 0.18 & 0.13 & 0.47 & 0.19 & 0.14 & 0.44   \\
B5R   &  0.6  &  0.08 & 0.12 & 0.10 & 0.46 & 0.12 & 0.09 & 0.52 & 0.13 & 0.12 & 0.42   \\
\hline
\multicolumn{12}{c} {Equipartition MHD Model} \\
\hline
E     & 10.0  &   5   & 0.73 & 0.21 & 0.46 & 0.81 & 0.25 & 0.43 & 1.90 & 0.49 & 0.22   \\
\hline
\multicolumn{12}{c}{Stochastic Models} \\
\hline
S2    & 10.0  &  20   & 1.37 & 0.15 & 0.52 & 1.29 & 0.15 & 0.53 & 1.31 & 0.14 & 0.53   \\  
S4    & 10.0  &  20   & 1.23 & 0.16 & 0.47 & 1.17 & 0.15 & 0.49 & 1.17 & 0.17 & 0.46   \\
\hline
\end{tabular}
\caption{}
\label{T3}
\end{table}

\clearpage

\clearpage

%

\clearpage

\clearpage

\clearpage

\clearpage

\clearpage

\clearpage

\end{document}